\documentclass[reprint,amsmath,amssymb,aps,floatfix]{revtex4-2}
\usepackage[utf8]{inputenc} 
\usepackage[T1]{fontenc}    
\usepackage[pdfusetitle]{hyperref}
\usepackage{graphicx,url,qrcode,bm,tikz,physics,dsfont,slashed,mathtools,booktabs}
\usepackage{subfig}
\usetikzlibrary{calc}
\usepackage{cleveref}
\usepackage{comment}
\crefname{section}{Sec.}{Secs.}
\Crefname{section}{Section}{Sections}
\crefname{figure}{Fig.}{Figs.}
\crefname{equation}{Eq.}{Eqs.}
\crefname{appendix}{Appendix}{Appendices}
\newcommand{\im}[0]{\mathrm{i}}
\DeclareMathOperator\eu{e}
\renewcommand{\vec}[1]{\bm{#1}}
\renewcommand{\Tr}[1]{\operatorname{Tr}\left\{#1\right\}}
\newcommand{\Heavi}[0]{\Theta }
\newcommand{\id}[0]{\mathds 1}
\newcommand{\diagram}[1]{\vcenter{\hbox{\begin{tikzpicture} #1 \end{tikzpicture}}}}

\newcommand{\dTwo}[0]{1.8cm}
\newcommand{\dThree}[0]{2.7cm}

\newcommand{\len}[0]{1.0}
\usepackage{tikz}
\usetikzlibrary{calc}
\tikzset{
  gluon/.style={thick, decorate, decoration={coil, amplitude=.08cm, segment length=.15cm, aspect=.75}}}
\tikzset{pion/.style={
	dashed,
	postaction={decorate},
	decoration={markings,mark=at position .6 with {\arrow{>}}}
	}}
\tikzset{nucleon/.style={
	postaction={decorate},
	decoration={markings,mark=at position .6 with {\arrow{>}}}
	}}
\tikzset{inmedium/.style={
	very thick,
	}}
\tikzset{source/.style={
	decorate,
	decoration={snake,amplitude=0.8mm}
	}}
\newcommand{\vertex}[2]{
	\fill[black!5] (#1) circle (.2);
	\draw (#1) circle (.2);
	\node at (#1) {\scalebox{0.8}{$#2$}};
}

\tikzset{>=latex}
\usetikzlibrary{datavisualization}
\usetikzlibrary{datavisualization.formats.functions}
\usetikzlibrary{patterns}    	             
\usetikzlibrary{decorations.markings}        
\usetikzlibrary{shapes.geometric}            
\usetikzlibrary{decorations.pathmorphing}    
\usetikzlibrary{decorations}

\newcommand{\SelfEnergyTypeTwo}[0]{
  \coordinate (A) at (0,0);
  \coordinate (B) at ($(A)+(\len,0)$);
  \coordinate (C) at ($(B)+(\len,0)$);
  \coordinate (D) at ($(B)+(0,-0.6)$);
  \draw[pion] (A) -- (B);
  \draw[pion] (B) -- (C);
  \draw[inmedium] ($(B)!0.5!(D)$) circle (.3cm);
  \node at (D) {\scalebox{0.7}{$\blacktriangleright$}};
}

\newcommand{\SelfEnergyTypeThree}[0]{
  \coordinate (A) at (0,0);
  \coordinate (B) at ($(A) + (\len,0)$);
  \coordinate (C) at ($(B) + (\len,0)$);
  \coordinate (E) at ($(B) + (-\len*0.7,-\len)$);
  \coordinate (F) at ($(B) + (\len*0.7,-\len)$);
  \draw[pion] (A) -- (B);
  \draw[pion] (B) -- (C);
  \draw[pion] (B) to[in=90,out=220] (E);
  \draw[pion] (F) to[in=320,out=90] (B);
  \draw[inmedium] (E) to[in=135,out=45] (F);
  \draw[inmedium] (E) to[in=225,out=-45] (F);
  \node at ($(E)!0.5!(F) + (0,0.3)$) {\scalebox{0.7}{$\blacktriangleright$}};
  \node at ($(E)!0.5!(F) + (0,-0.3)$) {\scalebox{0.7}{$\blacktriangleleft$}};
}

\newcommand{\SelfEnergyTypeFour}[0]{
  \coordinate (A) at (0,0);
  \coordinate (B) at ($(A)+(\len,0)$);
  \coordinate (C) at ($(B)+(\len,0)$);
  \coordinate (D) at ($(C)+(\len,0)$);
  \draw[pion] (A) -- (B);
  \draw[pion] (B) -- (C);
  \draw[pion] (C) -- (D);
  \draw[inmedium] ($(B)!0.5!(C)$) circle (.5cm);
  \node at ($(B)!0.5!(C) + (0,0.5)$) {\scalebox{0.7}{$\blacktriangleright$}};
  \node at ($(B)!0.5!(C) - (0,0.5)$) {\scalebox{0.7}{$\blacktriangleleft$}};
}

\newcommand{\SelfEnergyTypeFive}[0]{
  \coordinate (A) at (0,0);
  \coordinate (B) at ($(A)+(\len,0)$);
  \coordinate (C) at ($(B)+(\len,0)$);
  \coordinate (D) at ($(B)+(0,-\len)$);
  \draw[pion] (A) -- (B);
  \draw[pion] (B) -- (C);
  \draw[pion] (D) -- (B);
  \draw[inmedium] ($(B)!0.5!(D)$) circle (\len/2);
  \node at ($(B)!0.5!(D) + (\len/2,0)$) {\scalebox{0.7}{$\blacktriangledown$}};
  \node at ($(B)!0.5!(D) - (\len/2,0)$) {\scalebox{0.7}{$\blacktriangle$}};
}

\newcommand{\SelfEnergyFirstA}[0]{
  \coordinate (A) at (0,0);
  \coordinate (B) at ($(A)+(\len,0)$);
  \coordinate (C) at ($(B)+(\len,0)$);
  \coordinate (D) at ($(C)+(\len,0)$);
  \draw[pion] (A) -- (B);
  \draw[pion] (C) -- (D);
  \draw ($(B)!0.5!(C)$) ++(180:.5) arc (180:0:.5);
  \draw[inmedium] ($(B)!0.5!(C)$) ++(180:.5) arc (180:360:.5);
  \node at ($(B)!0.5!(C)+(0,.5)$) {\scalebox{0.7}{$\blacktriangleright$}};
  \node at ($(B)!0.5!(C)-(0,.5)$) {\scalebox{0.7}{$\blacktriangleleft$}};
  \vertex{B}{1};
  \vertex{C}{1};
}

\newcommand{\SelfEnergyFirstB}[0]{
  \coordinate (A) at (0,0);
  \coordinate (B) at ($(A)+(\len,0)$);
  \coordinate (C) at ($(B)+(\len,0)$);
  \coordinate (D) at ($(C)+(\len,0)$);
  \draw[pion] (A) -- (B);
  \draw[pion] (C) -- (D);
  \draw[inmedium] ($(B)!0.5!(C)$) ++(180:.5) arc (180:0:.5);
  \draw ($(B)!0.5!(C)$) ++(180:.5) arc (180:360:.5);
  \node at ($(B)!0.5!(C)+(0,.5)$) {\scalebox{0.7}{$\blacktriangleright$}};
  \node at ($(B)!0.5!(C)-(0,.5)$) {\scalebox{0.7}{$\blacktriangleleft$}};
  \vertex{B}{1};
  \vertex{C}{1};
}

\newcommand{\DecayTypeTwo}[0]{
  \coordinate (A) at (0,0);
  \coordinate (B) at ($(A)+(\len,0)$);
  \coordinate (C) at ($(B)+(\len,0)$);
  \coordinate (D) at ($(B)+(0,-0.6)$);
  \draw[pion] (A) -- (B);
  \draw[source] (B) -- (C);
  \draw[inmedium] ($(B)!0.5!(D)$) circle (.3cm);
  \node at (D) {\scalebox{0.7}{$\blacktriangleright$}};
}

\newcommand{\DecayTypeThree}[0]{
  \coordinate (A) at (0,0);
  \coordinate (B) at ($(A) + (\len,0)$);
  \coordinate (C) at ($(B) + (\len,0)$);
  \coordinate (E) at ($(B) + (-\len*0.7,-\len)$);
  \coordinate (F) at ($(B) + (\len*0.7,-\len)$);
  \draw[pion] (A) -- (B);
  \draw[source] (B) -- (C);
  \draw[pion] (B) to[in=90,out=220] (E);
  \draw[pion] (F) to[in=320,out=90] (B);
  \draw[inmedium] (E) to[in=135,out=45] (F);
  \draw[inmedium] (E) to[in=225,out=-45] (F);
  \node at ($(E)!0.5!(F) + (0,0.3)$) {\scalebox{0.7}{$\blacktriangleright$}};
  \node at ($(E)!0.5!(F) + (0,-0.3)$) {\scalebox{0.7}{$\blacktriangleleft$}};
}

\newcommand{\DecayTypeFour}[0]{
  \coordinate (A) at (0,0);
  \coordinate (B) at ($(A)+(\len,0)$);
  \coordinate (C) at ($(B)+(\len,0)$);
  \coordinate (D) at ($(C)+(\len,0)$);
  \draw[pion] (A) -- (B);
  \draw[pion] (B) -- (C);
  \draw[source] (C) -- (D);
  \draw[inmedium] ($(B)!0.5!(C)$) circle (.5cm);
  \node at ($(B)!0.5!(C) + (0,0.5)$) {\scalebox{0.7}{$\blacktriangleright$}};
  \node at ($(B)!0.5!(C) - (0,0.5)$) {\scalebox{0.7}{$\blacktriangleleft$}};
}

\newcommand{\DecayTypeFive}[0]{
  \coordinate (A) at (0,0);
  \coordinate (B) at ($(A)+(\len,0)$);
  \coordinate (C) at ($(B)+(\len,0)$);
  \coordinate (D) at ($(B)+(0,-\len)$);
  \draw[pion] (A) -- (B);
  \draw[source] (B) -- (C);
  \draw[pion] (D) -- (B);
  \draw[inmedium] ($(B)!0.5!(D)$) circle (\len/2);
  \node at ($(B)!0.5!(D) + (\len/2,0)$) {\scalebox{0.7}{$\blacktriangledown$}};
  \node at ($(B)!0.5!(D) - (\len/2,0)$) {\scalebox{0.7}{$\blacktriangle$}};
}

\newcommand{\DecayConstantFirstA}[0]{
  \coordinate (A) at (0,0);
  \coordinate (B) at ($(A)+(\len,0)$);
  \coordinate (C) at ($(B)+(\len,0)$);
  \coordinate (D) at ($(C)+(\len,0)$);
  \draw[pion] (A) -- (B);
  \draw[source] (C) -- (D);
  \draw ($(B)!0.5!(C)$) ++(180:.5) arc (180:0:.5);
  \draw[inmedium] ($(B)!0.5!(C)$) ++(180:.5) arc (180:360:.5);
  \node at ($(B)!0.5!(C)+(0,.5)$) {\scalebox{0.7}{$\blacktriangleright$}};
  \node at ($(B)!0.5!(C)-(0,.5)$) {\scalebox{0.7}{$\blacktriangleleft$}};
  \vertex{B}{1};
  \vertex{C}{1};
}

\newcommand{\DecayConstantFirstB}[0]{
  \coordinate (A) at (0,0);
  \coordinate (B) at ($(A)+(\len,0)$);
  \coordinate (C) at ($(B)+(\len,0)$);
  \coordinate (D) at ($(C)+(\len,0)$);
  \draw[pion] (A) -- (B);
  \draw[source] (C) -- (D);
  \draw[inmedium] ($(B)!0.5!(C)$) ++(180:.5) arc (180:0:.5);
  \draw ($(B)!0.5!(C)$) ++(180:.5) arc (180:360:.5);
  \node at ($(B)!0.5!(C)+(0,.5)$) {\scalebox{0.7}{$\blacktriangleright$}};
  \node at ($(B)!0.5!(C)-(0,.5)$) {\scalebox{0.7}{$\blacktriangleleft$}};
  \vertex{B}{1};
  \vertex{C}{1};
}

\begin{document}
\title{Pion properties in isospin-asymmetric nuclear matter using in-medium chiral perturbation theory}
\author{Kihong Kwon}
\email{kwon.k.297a@m.isct.ac.jp}
\author{Yamato Suda}
\email{suda.y.bf53@m.isct.ac.jp}
\author{Stephan Hübsch}
\author{Daisuke Jido}
\email{jido@phys.titech.ac.jp}
\affiliation{Department of Physics, Institute of Science Tokyo, Meguro, Tokyo 152-8551, Japan}
\date{\today}
\begin{abstract}
We compute the density dependence of in-medium pion properties, such as mass, wave function renormalization, and decay constant in the correlation function approach, and how they change under the influence of isospin-asymmetric nuclear matter. 
To this end, we use in-medium chiral perturbation theory to compute the relevant Feynman diagrams up to two-loop diagrams. 
Our results show that the isospin asymmetry of the nuclear matter splits these quantities into three separate values, corresponding to the three pions. Consequently, the tendency of each in-medium pion mass, wave function renormalization, and decay constant is dependent on the density and the neutron-to-proton ratio $\rho_n/\rho_p$ of nuclear matter. We also derive an in-medium Gell-Mann--Oakes--Renner relation which is valid for isospin-asymmetric nuclear matter and investigate to what extent it holds within our calculations. 
\end{abstract}
\maketitle

\section{Introduction}
The low-energy spectrum of quantum chromodynamics (QCD) is heavily influenced by the dynamical breaking of chiral symmetry (DB$\chi$S). 
Recent experimental results~\cite{Suzuki2004,Friedman2004,nishi2023chiral,Friedman2005,nishi2018spectroscopy,geissel2002deeply,Itahashi:1999qb,yamazaki1998effective,friedman2002indications,grion2005pi,palmese2016chiral,shimizu2003search,larionov2022effects,friedman2001evidence,suzuki2003observation,sorge1997psi,ejima2025toward,moreau2017evidence,sung2024k}, as well as theoretical considerations based on these results~\cite{ikeno2011precision,ikeno2023pion,tani2021structure,Kolomeitsev2003,Jido2008,birse1994chiral,hatsuda1999precursor,celenza1992partial,birse1994two,jido2007medium,marczenko2022chiral,jido2013eta,hirenzaki1991d,toki1989structure,galain1991vector,fernandez2007pion,suzuki2016d,garcia2002chiral,gubler2017novel,chanfray2003scalar,jin2022quark,bernard1997chiral,narayanan2006chiral,meisinger1996chiral,glozman2024chiral,lenaghan2000chiral}, have revealed a partial restoration of chiral symmetry in nuclear matter and at finite temperature. 
Since DB$\chi$S is supposed to be the origin of most of the mass in our universe, one can expect hadronic properties to change following the partial restoration of chiral symmetry. 
In this work, we examine the pions, which are the lightest hadrons due to their pseudo-Goldstone nature, and how their properties change in nuclear matter. 

This work is a continuation of an earlier study~\cite{Goda2014}, where such pion properties were calculated in isospin-symmetric nuclear matter. 
The main distinction of the present work is that we consider the more general case of isospin-asymmetric nuclear matter, where the densities of protons and neutrons can be varied independently. 
Due to the effect of isospin asymmetry of nuclear matter, each pion employs distinct properties according to its isospin. To separate each result, we utilize the physical basis of the pions. 
In the present work, we also suggest that more diagrams are necessary to fully describe the pion properties. 
This includes topologically similar diagrams where different interaction vertices are used, as well as topologically distinct diagrams, which become especially important in isospin-asymmetric nuclear matter. 

There have been several other works investigating in-medium pion properties after Refs.~\cite{Oller2002,Meisner2002} introduced in-medium chiral perturbation theory. 
The in-medium wave function renormalization has been investigated by Ref.~\cite{Chanfray2003}, where chiral perturbation theory approach was connected to a linear sigma model calculation, and by Ref.~\cite{Kolomeitsev2003}, using two-loop chiral perturbation theory in order to calculate the pion-nuclear $s$-wave optical potential. 
The in-medium pion masses were studied in Ref.~\cite{Kaiser2001} by using two-loop chiral perturbation theory, in Refs.~\cite{Drews2015,Drews2017} using functional renormalization group methods, and also in Ref.~\cite{Meisner2002} which used one-loop in-medium chiral perturbation theory. 
The in-medium decay constant has been investigated in Refs.~\cite{Meisner2002,Lacour2010} by the chiral perturbation theory in nuclear matter, and in Ref.~\cite{Hutauruk2019} with an NJL model. 
On the experimental side, Ref.~\cite{Itahashi2000} investigated pionic lead atoms and determined in-medium parameters of the pion-nucleon optical potential, which was then used to determine the in-medium mass increase of the negatively charged pion. 
A later study~\cite{Suzuki2004} investigated pionic tin atoms and was able to determine the in-medium corrections to the $b_1^*$ parameter of the $s$-wave pion-nucleus potential. 
Furthermore, Ref.~\cite{Kienle2004} used 1$s$-state binding energies and widths of pionic atoms to derive the $s$-wave pion-nucleus potential parameters and connected this to an in-medium decrease of the $\pi^-$ decay constant. 
Let us also briefly mention other related works that deal with nuclear matters and related hadronic effects. 
Refs.~\cite{Khunjua2018,Khunjua2018a,Khunjua2018b,Khunjua2019,buballa2005njl,xia2024extended,wang2019quark,ratti2003njl,alaverdyan2022quark,mishra2022colorsuperconductivity,liu2016isospin,wang2020stability,khunjua2019chiral} and related works investigated quark matters and the possibility of pion condensation in the NJL model. 
Further meson condensation processes were investigated by Refs.~\cite{Mammarella2016,Mammarella2017}, which considered both isospin-asymmetric nuclear matter and strange nuclear matter. 
In Ref.~\cite{Kolomeitsev2003}, the authors calculate the pion-nuclear $s$-wave optical potential in isospin-asymmetric nuclear matter. 
Ref.~\cite{Li2017} investigated $\Delta$ formation cross section and its decay width in isospin-asymmetric nuclear matter and Ref.~\cite{Kovensky2021} investigated isospin-asymmetric baryonic matter in the  holographic Witten--Sakai--Sugimoto model. 
Finally, we note that by using the formalism described in this work, it is also possible to calculate the in-medium corrections to the quark condensate via a different set of Feynman diagrams~\cite{Goda2013,Huebsch2021}. 

The structure of this paper is as follows:
In \cref{sec:in-medium-pion-properties}, we review the in-medium pion states and show how to compute the in-medium pion properties using the correlation function approach. In \cref{sec:in-medium-GOR}, we derive the Gell-Mann-Oakes-Renner relation in isospin-asymmetric nuclear matter.
In \cref{sec:methods}, we review the in-medium chiral perturbation theory and a chiral Lagrangian, which we use to do our calculations.
In \cref{sec:results} we show the Feynman diagrams that we considered in this work as well as our numerical results for the in-medium mass, wave function renormalization and decay constant of each pion, and finally \cref{sec:summary} is devoted to the summary. 

\section{Pion properties in asymmetric nuclear matter}
\label{sec:in-medium-pion-properties}

In this section, we define the in-medium pion properties, such as the in-medium pion mass $m_\pi^*$, the wave function renormalization $Z_\pi$, and the in-medium coupling constants $G_\pi^*$ and $f^*_\pi$ as matrix elements of the pseudoscalar and axial-vector currents for the in-medium pion states. They will be accessed as the properties of the pion pole in the in-medium correlation functions of these currents, which can be calculated by using the in-medium chiral perturbation theory that we will formulate in \cref{sec:methods}.

\subsection{In-medium pion properties}
 
We define the in-medium pion state by following Ref.~\cite{Goda2014}.  The pion states, $|\pi\rangle$ in vacuum and $|\pi^* \rangle$ in medium, are normalized as
\begin{subequations}
\begin{align}
    \langle\pi^{\pm,0} (k_{\pi^{\pm,0}}) | \pi^{\pm,0}(p_{\pi^{\pm,0}})\rangle &= (2\pi)^3 2E_{\pi^{\pm,0}}(\vec p) \delta^{(3)}(\vec p - \vec k), \\ 
    \langle\pi^{*\pm,0} (k_{\pi^{\pm,0}}) | \pi^{*\pm,0}(p_{\pi^{\pm,0}})\rangle &= (2\pi)^3 2\omega_{\pi^{\pm,0}}(\vec p) \delta^{(3)}(\vec p - \vec k), \label{eq:states_norm}
\end{align}
\end{subequations}
where we have expressed the charged and neutral pions collectively as $\pi$ or $\pi^{\pm,0}$ and the momenta of the pions are measured in the nuclear matter rest frame, and the in-medium four-momentum of an on-shell pion is written as $p_\pi^\mu=(\omega_\pi(\vec p),\vec p)$.
The energy $E_{\pi}(\vec p)$ for the in-vacuum pions satisfies simply the in-vacuum dispersion relation $E_{\pi}(\vec p)^2 = m_\pi^2 + \vec{p}^2$ with the in-vacuum pion mass $m_\pi$, while the in-medium energy $\omega_{\pi}(\vec p)$ is not such a simple relation due to the medium effects and its form is obtained as the pole position of the correlation function that is explained later. Note that $\omega_\pi(\vec p)=\omega_\pi(-\vec p)$ in isotropic matter. The in-medium pion mass $m_{\pi}^{*}$ is defined by the in-medium pion energy at rest as 
\begin{align}
m_{\pi^{\pm,0}}^{*} = \omega_{\pi^{\pm,0}}(\vec 0).  \label{eq:def-m}
\end{align}

Giving the normalization of the pion fields by the in-vacuum state as 
\begin{align}
    \langle0|\pi^{\pm,0}(x)|\pi^{\pm,0}(p_{\pi^{\pm,0}})\rangle = \eu^{-\im p_{\pi^{\pm,0}} \cdot x}, \label{eq:field_norm}
\end{align}
we introduce the matrix element of the pion field in a nuclear medium as 
\begin{align}
\langle\Omega|\pi^{\pm,0}(x)|\pi^{*\pm,0}(p_{\pi^{\pm,0}})\rangle &= \sqrt{Z_{\pi^{\pm,0}}} \eu^{-\im p_{\pi^{\pm,0}} \cdot x}, \label{eq:field_norm_medium}
\end{align}
with the wave function renormalization $Z_{\pi}$. 
Here the asymptotic state of the pure vacuum state is denoted by $|0\rangle$ and that of the nuclear matter is by $|\Omega\rangle$. In the same way, the matrix element of the pseudoscalar current $P$ in medium is written by 
\begin{align}
    \langle\Omega|P^{\pm,0}(x)|\pi^{*\pm,0}(p_{\pi^{\pm,0}})\rangle = G^*_{\pi^{\pm,0}}(p_{\pi^{\pm,0}}) \eu^{-\im p_{\pi^{\pm,0}} \cdot x}, \label{eq:P_coupling_medium}
\end{align}
with the in-medium form factor $G^*_\pi(p_{\pi^{\pm,0}})$, which satisfies $G^*_\pi(\omega_\pi,\vec p)=G^*_\pi(\omega_\pi,-\vec p)$ in isotropic nuclear matter. The pseudoscalar currents are defined as 
\begin{align}
    P^{\pm}= \frac 1 {\sqrt{2}}(P^1 \mp \im P^2), \qquad P^0=P^3,
\end{align}
in terms of the quark fields 
\begin{align}
   P^a = \bar q \im \gamma_5 \tau^a q,
\end{align}
with the Pauli matrix $\tau^a$ in the isospin space and the quark field $q = (u,d)^T$. For later convenience, we present the complex conjugate of \cref{eq:P_coupling_medium}:
\begin{align}
    \langle \pi^{*\pm,0}(p_{\pi^{\pm,0}}) |P^{\mp,0}(x)|\Omega\rangle =\bar{ G}_{\pi^{\pm,0}}^*(p_{\pi^{\pm,0}}) \eu^{\im p_{\pi^{\pm,0}} \cdot x}, \label{eq:P_coupling_medium_cc}
\end{align}
where $\bar{P}^\pm=P^\mp$ and the bar symbol means complex conjugate.

The decay constant in nuclear matter is defined as the matrix element of an axial-vector current $A_\mu^a$, that is one of the Noether currents of the SU(2) chiral symmetry, between a one-pion in-medium state and the nuclear matter state~\cite{Goda2014}.
Here we introduce the matrix elements in a slightly different form from Ref.~\cite{Jido2008}:
\begin{align}\label{eq:axi}
       &\langle\Omega|A_\mu^{\pm,0}(x)|\pi^{* \pm,0}(p_{\pi^{\pm,0}})\rangle\nonumber\\
       &\qquad=\im[n_\mu N_{\pi^{\pm,0}}^*(p_{\pi^{\pm,0}})+p^{\pi^{\pm,0}}_\mu F_{\pi^{\pm,0}}^*(p_{\pi^{\pm,0}})]\eu^{-\im p_{\pi^{\pm,0}}\cdot x},
\end{align} 
where the axial-vector currents $A^{\pm, 0}_{\mu}$ are defined by
\begin{align}
   A^{\pm}_{\mu} = \frac 1 {\sqrt 2} (A^{1}_{\mu} \mp \im A^{2}_{\mu}), \qquad A^{0}_{\mu} = A^{3}_{\mu},
\end{align}
which are given in terms of the quark fields as 
\begin{align}
    A^{a}_{\mu} = \bar q \gamma_{\mu} \gamma_{5} \frac{\tau^{a}}{2} q, 
\end{align}
and $n_\mu$ is a vector describing the Lorentz frame of the nuclear matter, which is $n_\mu=(1,0,0,0)$ in the rest frame. Note that $N_\pi^*$ has one more mass dimension than $F_\pi^*$, and can be decomposed to
\begin{align}\label{eq:ndec}
    N_\pi^*(p_{\pi})=N_{1\pi}^*(p_{\pi})+(n\cdot p_\pi)N_{2\pi}^*(p_\pi), 
\end{align}
where $N_{1\pi}^*(p_\pi)$ and $N_{2\pi}^*(p_\pi)$ are even functions of $\vec p$ in isotropic matter. We often drop the momentum dependence of the form factors for convenience. 
The temporal and spatial components of the matrix elements in the rest frame of nuclear matter are written as 
\begin{subequations}\label{eq:dedeco}
\begin{align}
	\langle\Omega|A_0^{\pm,0}(x)|\pi^{*\pm,0}(p_{\pi^{\pm,0}})\rangle &= \im (N^{*}_{\pi^{\pm,0}} +  F^{*}_{\pi^{\pm,0}} \omega_{\pi^{\pm,0}})\eu^{-\im p_{\pi^{\pm,0}}\cdot x},
\label{q:dedecoa}\\
	\langle\Omega|A_i^{\pm,0}(x)|\pi^{*\pm,0}(p_{\pi^{\pm,0}})\rangle &= \im F^{*}_{\pi^{\pm,0}} p_i \eu^{-\im p_{\pi^{\pm,0}}\cdot x}.
\label{q:dedecob}
\end{align}
\end{subequations}
If one defines the temporal and spatial components of the in-medium pion decay constant as
\begin{subequations}\label{eq:def-f}
\begin{align}
	\langle\Omega|A_0^{\pm,0}(x)|\pi^{*\pm,0}(p_{\pi^{\pm,0}})\rangle &= \im f^{*(t)}_{\pi^{\pm,0}} \omega_{\pi^{\pm,0}}\eu^{-\im p_{\pi^{\pm,0}}\cdot x}, \\
	\langle\Omega|A_i^{\pm,0}(x)|\pi^{*\pm,0}(p_{\pi^{\pm,0}})\rangle &= \im f^{*(s)}_{\pi^{\pm,0}} p_i\eu^{-\im p_{\pi^{\pm,0}}\cdot x},
\end{align}
\end{subequations}
the decay constants have the following relations at the pion rest frame:
\begin{align}
    f_\pi^{*(t)}=F_\pi^*+\frac{N_\pi^*}{m^{*}_{\pi}},\qquad f_\pi^{*(s)}=F_\pi^*.
\label{eq:chfn}
\end{align}
This expression should be carefully evaluated in the chiral limit for the massless mode because the asymmetry nuclear matter breaks chiral symmetry explicitly and the matrix elements \eqref{eq:def-f} may not be proportional to the pion four-momentum $p_{\mu}$.

\subsection{Correlation function approach}\label{sec:cfa}
Let us see how to calculate the in-medium properties of the pions. We make good use of correlation functions in a nuclear medium, and the pion properties are expressed by the pole structures of the correlation functions. Let us consider the correlation function of the pseudoscalar currents defined as 
\begin{align}
\Pi^{\pm\mp,00}(p) 
& \equiv \int \dd[4] x \eu^{\im p\cdot x} \langle\Omega|\mathsf T P^{\pm,0} (x) P^{\mp,0}(0)|\Omega\rangle.
\label{eq:def_PP_corr}
\end{align}
In order to pick up the pion contributions from the correlation function \eqref{eq:def_PP_corr}, we insert the complete set of the in-medium pion states and use the matrix elements \eqref{eq:P_coupling_medium} and \eqref{eq:P_coupling_medium_cc}.
The detailed calculations are demonstrated in \cref{app:MediumGORPhysBasis}.
Finally we find
\begin{align}
    &\ \Pi^{\pm \mp,00}(p) \nonumber \\ 
    &= \frac{1}{2 \omega_{\pi^{\pm,0}}(\vec p)} G^*_{\pi^{\pm,0}} \frac{\im}{p_0 - \omega_{\pi^{\pm,0}}(\vec p) + \im \epsilon} \bar{G}^{*}_{\pi^{\pm,0}}\nonumber \\ 
    &\ \quad - \frac{1}{2 \omega_{\pi^{\mp,0}}(\vec p)} G^*_{\pi^{\mp,0}} \frac{\im}{p_0 + \omega_{\pi^{\mp,0}}(\vec p) - \im \epsilon} \bar{G}^{*}_{\pi^{\mp,0}} 
    + \ldots,
    \label{eq:PP_corr_inmedium}
\end{align}
where the positive and negative pole positions for the charged pions can be different, $\omega_{\pi^{\pm}}(\vec p) \neq \omega_{\pi^{\mp}}(\vec p)$,  due to medium effects that break the isospin symmetry. In general, the correlation function should have other contributions than those coming from pion states. The pion properties can be extracted from the pion pole position and its residue. For the complex conjugate, one can obtain the relation
\begin{align}\label{eq:ppcp}
    &\big[\Pi^{\pm\mp,00}(p_0,\vec p)\big]^\dagger\nonumber\\
    &=-\frac{1}{2 \omega_{\pi^{\pm,0}}(\vec p)} G^*_{\pi^{\pm,0}} \frac{\im}{p_0 - \omega_{\pi^{\pm,0}}(\vec p) - \im \epsilon} \bar{G}^{*}_{\pi^{\pm,0}}\nonumber \\ 
    &\ \quad + \frac{1}{2 \omega_{\pi^{\mp,0}}(\vec p)} G^*_{\pi^{\mp,0}} \frac{\im}{p_0 + \omega_{\pi^{\mp,0}}(\vec p) + \im \epsilon} \bar{G}^{*}_{\pi^{\mp,0}} 
    + \ldots\nonumber\\
    &=-\Pi^{\mp\pm,00}(-p_0,\vec p).
\end{align}

In this paper, we consider the case away from the chiral limit where the in-vacuum pion mass is non-zero.
The correlation function \eqref{eq:def_PP_corr} can be expressed by using the pion self-energy $\Sigma_{\pi}$ and the vertex correction $\hat G_{\pi}$ as 
\begin{align}
   \Pi^{\pm\mp,00}(p) = 
    \hat G_{\pi^{\pm,0}} 
    \frac{\im}{p^2-m_\pi^2-\Sigma^{\pm,0}_\pi(p) + \im \epsilon}
    \bar{\hat G}_{\pi^{\pm,0}} + \cdots.
     \label{eq:p0p0-vac}
\end{align}
All of the medium effects are summarized into the self-energy $\Sigma_{\pi}$ and the vertex correction $\hat G_{\pi}$, which is given by one-particle irreducible diagrams. In order to separate out the positive energy solution, we decompose the in-medium pion propagator as  
\begin{align}\label{eq:vapp}
    &\ \Pi^{\pm\mp,00}(p) \nonumber \\ 
    &=\hat G_{\pi^{\pm,0}} \frac{\im}{2\mathcal{E}_{\pi^{\pm,0}}}\left(
    \frac{1}{p_0-\mathcal{E}_{\pi^{\pm,0}} + \im \epsilon}-\frac{1}{p_0+\mathcal{E}_{\pi^{\pm,0}} - \im \epsilon}\right)
    \bar{\hat G}_{\pi^{\pm,0}}\nonumber\\
    &\quad+\ldots,
\end{align}
where $\mathcal{E}_\pi=\sqrt{m_\pi^2+\Sigma_\pi(p)+\vec p^2}$. The pole position can be found by solving $p_0-\mathcal{E}_\pi=0$ for $p_0$ with a fixed $\vec p$. The solution of the equation,
\begin{align}\label{eq:mass}
   \omega_{\pi^{\pm,0}}(\vec p) = \sqrt{m^{2}_{\pi} + \Sigma^{\pm,0}_{\pi}( \omega_{\pi^{\pm,0}}(\vec p), \vec p) + \vec p^{2}},
\end{align}
gives the dispersion relation of the in-medium pion. 
Because the in-medium mass is given by \cref{eq:def-m}, it is calculated by solving the self-consistent equation  
\begin{align}
   m^{*}_{\pi^{\pm,0}} = \sqrt{m^{2}_{\pi} + \Sigma^{\pm,0}_{\pi}( m^{*}_{\pi^{\pm,0}}, \vec 0)}. \label{eq:pionmass}
\end{align}
Introducing an in-medium kinetic energy as
\begin{align}
  {\mathcal K}_{\pi}(\vec p)  \equiv \omega_{\pi} (\vec p) - m_{\pi}^{*},
\end{align}
we write the pole position for the positive energy solution as
\begin{align}
p_0-\mathcal{E}_\pi = Z_{\pi}^{-1} (p_0 - m_{\pi}^{*} - \mathcal{K}_{\pi}) + \cdots,
\end{align}
where $\cdots$ means contributions with higher orders of $(p_0 - m_{\pi}^{*} - \mathcal{K}_{\pi})$ and 
the wave function renormalization $Z_{\pi}$ can be calculated by
\begin{align}
    Z_\pi =\left(1-\frac{1}{2\omega_\pi (\vec p) }\left.\frac{\partial\Sigma_\pi(p)}{\partial p_0}\right|_{p_{0}=\omega_\pi (\vec p) }\right)^{-1}\label{eq:def-Z}.
\end{align}
In this way,
we write the correlation function \eqref{eq:p0p0-vac} for the positive energy solution as 
\begin{align}
    &\  \Pi^{\pm\mp,00}(p) \nonumber \\ 
    &= \frac{1}{2 \omega_{\pi^{\pm,0}}} 
    \hat G_{\pi^{\pm,0}} 
    \frac{\im Z_{\pi^{\pm,0}}}{p_0 - m^*_{\pi^{\pm,0}}-\mathcal{K}_{\pi^{\pm,0}}(\vec p) + \im \epsilon}
    \bar{\hat G}_{\pi^{\pm,0}}+ \ldots.
     \label{eq:invacuum_corre}
\end{align}
Comparing \cref{eq:PP_corr_inmedium,eq:invacuum_corre}, we obtain 
\begin{align}
    G^*_{\pi^{\pm,0}} &= \hat G_{\pi^{\pm,0}} \sqrt{Z_{\pi^{\pm,0}}} \label{eq:G*_G_sqrtZ},
\end{align}
at the pion pole $p_{0}= \omega_{\pi}(\vec p)$. The importance to take into account of the wave function renormalization in the in-medium quantities was pointed out in Refs. \cite{Jido2008,hatsuda1999precursor}. Note that the negative energy solution of \cref{eq:mass} corresponds to the antiparticle of a field. Thus, self-energies that are even functions of $\vec p$ in isotropic matter have the relation under $p_0\leftrightarrow-p_0$:
\begin{align}\label{eq:ses}
    \Sigma_{\pi}^{\pm,0}(-p_0,\vec p)=\Sigma_{\pi}^{\mp,0}(p_0,\vec p).
\end{align}
Then, one can find the following relation for the vertex correction $\hat G_\pi$ from \cref{eq:ppcp,eq:vapp,eq:ses}:
\begin{align}\label{eq:gsign}
    \hat G_{\pi^{\pm,0}}(-p_0,\vec p)=\hat G_{\pi^{\mp,0}}(p_0,\vec p).
\end{align}
We next consider the correlation function of the axial-vector currents defined as
\begin{align}
    \Pi^{\pm \mp,00}_{\mu\nu}(p) \equiv \int \dd[4] x \eu^{\im p\cdot x} \langle\Omega|\mathsf{T}A_\mu^{\pm,0}(x)A_\nu^{\mp,0}(0)|\Omega\rangle. \label{eq:def_AA_corr}
\end{align}
This can be written in terms of the in-medium quantities, as derived in \cref{app:MediumGORPhysBasis}:
\begin{widetext}
\begin{align}\label{eq:aacor_tmp}
    \Pi^{\pm\mp,00}_{\mu\nu}(p) 
    &= \im\frac{(n_\mu N_{\pi^{\pm,0}}^*+p_\mu^{\pi^{\pm,0}}(\vec p) F^*_{\pi^{\pm,0}})(n_\nu \bar N_{\pi^{\pm,0}}^*+p_\nu^{\pi^{\pm,0}}(\vec p) \bar F^*_{\pi_\pi^{\pm,0}})}{2\omega_{\pi^{\pm,0}}(\vec p)\big(p^0 -\omega_{\pi^\pm}(\vec p) + \im \epsilon\big)}+\ldots,\nonumber\\
\end{align}
\end{widetext}
for the positive energy pole. Similarly to the correlation function of the pseudoscalar currents, we obtain for $\mu=\nu$
\begin{align}
    \big[\Pi^{\pm\mp,00}_{\mu\mu}(p_0,\vec p)\big]^\dagger=-\Pi^{\mp\pm,00}_{\mu\mu}(-p_0,\vec p),
\end{align}
where $\mu$ is not summed over. The correlation function (\ref{eq:def_AA_corr}) can be written by using the self-energy $\Sigma_\pi$ and the vertex correction $\im\hat \Gamma_\mu^\pi$ as
\begin{widetext}
\begin{align}
    \Pi^{\pm\mp,00}_{\mu\nu}(p) 
    = 
    (\im\hat \Gamma_\mu^{\pi^{\pm,0}}) 
    \frac{\im }{p^2-m_\pi^2-\Sigma_\pi^{\pm,0}(p) + \im \epsilon}
    (-\im\bar{\hat \Gamma}_\nu^{\pi^{\pm,0}})+\ldots.
    \label{eq:invacuumaa}
\end{align}
Expanding this correlation function around the in-medium pion pole, i.e., $p_0 = \omega_{\pi^{\pm,0}}(\vec p)$, we find
\begin{align}
    \Pi^{\pm\mp,00}_{\mu\nu}(p) 
    = 
    \hat \Gamma_\mu^{\pi^{\pm,0}}
    \frac{\im Z_{\pi^{\pm,0}} }{2\omega_{\pi^{\pm,0}}(\vec p)\big(p_0 - m^*_{\pi^{\pm,0}}-\mathcal{K}_{\pi^{\pm,0}}(\vec p) + \im \epsilon\big)} \bar{\hat \Gamma}_\nu^{\pi^{\pm,0}} + \ldots. \label{eq:aacor_medium_tmp}
\end{align}
\end{widetext}
By comparing (\ref{eq:aacor_tmp}) and (\ref{eq:aacor_medium_tmp}), we can calculate $F_\pi^*$ and $N_\pi^*$ as
\begin{align}
    n_\mu N_{\pi}^*+p_\mu^{\pi}(\vec p) F^*_{\pi} = \hat \Gamma_\mu^{\pi} \sqrt{Z_{\pi}}.
\label{eq:rede}
\end{align}
Note that the wave function renormalization plays an important role in taking into account the effect of partial restoration of chiral symmetry at finite density to the decay constants correctly \cite{Jido2008,hatsuda1999precursor}. Here, the vertex correction can be calculated by considering one-particle irreducible (1PI) diagram contributions, which can be calculated in the in-medium chiral perturbation theory. Then, we can calculate the decay constants using~\cref{eq:chfn}. 
When one has a nonzero pion mass, we define an effective vertex correction as
\begin{align}\label{eq:effvex}
    p_\mu\hat f_\pi\equiv\hat \Gamma_\mu^\pi. 
\end{align}    
Then, the temporal component of the decay constant is calculated in the rest frame of the pion as
\begin{align}\label{eq:fd}
    f_\pi^{*(t)}=\sqrt{Z_\pi}\hat f_\pi.
\end{align}
We often omit the index $t$ for this quantity, and call it decay constant. Note that we have the symmetry under $\vec p\leftrightarrow-\vec p$ for the form factors $N_\pi^*(p_\pi)$, $F_\pi^*(p_\pi)$, and $G_\pi^*(p_\pi)$ in isotropic nuclear matter, and the coupling constants $N_\pi^*$, $F_\pi^*$, and $G_\pi^*$ are calculated from the form factors at the pole position $p^\mu=(m_\pi^*,\vec 0)$ for the rest pion.

In summary, after computing the pion self-energies using the in-medium chiral perturbation theory, we obtain the in-medium pion masses, wave function renormalizations, and coupling constants using \cref{eq:pionmass,eq:def-Z,eq:G*_G_sqrtZ,eq:rede,eq:fd}.

\section{In-medium Gell-Mann--Oakes--Renner relation}\label{sec:in-medium-GOR}
The Gell-Mann--Oakes--Renner (GOR) relation~\cite{Gell-Mann1968} connects the quark condensate $\langle \bar qq\rangle = \langle \bar uu+\bar dd\rangle$ to hadronic (e.g., pion) quantities. In vacuum, it is given as 
\begin{align}
	m_\pi^2 f_\pi^2 &= -\frac{m_u + m_d}{2} \langle \bar uu + \bar dd \rangle =-m_q \langle \bar qq\rangle,
\end{align}
where $m_u$ and $m_d$ are the up and down quark masses. In order to derive an in-medium version of the GOR relation, we follow Refs.~\cite{Jido2008,Hubsch2021}, but generalize their discussion to isospin-asymmetric nuclear matter. We consider the following correlation functions, 
\begin{align}
    \Pi^{\pm\mp,00}_5(q) \equiv \int \dd[4] x \eu^{\im q\cdot x} \partial^\mu \langle\Omega|\mathsf T A_\mu^{\pm,0} (x) P^{\mp,0}(0)|\Omega\rangle.
\end{align}
Using the partially conserved axial current (PCAC) relation $\partial^\mu A_\mu^{\pm,0}(x)=m_qP^{\pm,0}(x)$, we can write this correlation function in terms of the in-medium quark condensate $\langle\bar qq\rangle^*$ in the limit of the rest frame, i.e., $\vec q \to \vec 0$ while keeping $q^0$ arbitrary as 
\begin{align}
     \lim_{\vec q \to 0}iq^\mu\Pi^{\pm\mp,00}_{5\mu}(q)&+m_q\lim_{\vec q\rightarrow0}\Pi^{\pm\mp,00}(q)\nonumber\\
    &=-\langle\Omega|[Q_5^{\pm,0},P^{\mp,0}(0)]|\Omega\rangle\nonumber\\
    &=\im\langle\bar qq\rangle^*,
\end{align}
where the correlation functions are defined by 
\begin{subequations}\label{eq:corr-function-phys}
\begin{align}
    \Pi^{\pm\mp,00} (q) &= \int \dd[4]x \eu^{\im q \cdot x} \langle \Omega | \mathsf T P^{\pm,0}(x) P^{\mp,0}(0)|\Omega \rangle, \\ 
    \Pi^{\pm\mp,00}_{5\mu} (q) &= \int \dd[4]x \eu^{\im q \cdot x} \langle \Omega | \mathsf T A_\mu^{\pm,0}(x) P^{\mp,0}(0)|\Omega \rangle.
\end{align}
\end{subequations}
Substituting the hadronic completeness and using the matrix elements of the axial-vector and pseudoscalar currents between one in-medium pion state and an asymptotic state in nuclear matter, we obtain the exact sum rule for all densities in the asymmetric nuclear matter using the partially conserved axial-vector current (PCAC) relation $\partial^\mu A_\mu^{\pm,0}=m_qP^{\pm,0}$ in the following form \cite{Jido2008}:
\begin{align}\label{eq:pcacim}
    (\omega_{n^{\pm,0}}(\vec p)^2- \vec p \cdot \vec p )F_{\pi_n^{\pm,0}}^{*}+\omega_{n^{\pm,0}}(\vec p)N_{\pi_n^{\pm,0}}^{*}  = m_q G_{\pi_n^{\pm,0}}^*.
\end{align}
Note that \Cref{eq:pcacim} is evaluated in the nuclear matter rest frame. The sum rule for the pion states at rest is given as:
\begin{widetext}
\begin{subequations}\label{eq:gorre}
\begin{align}
    \sum_{n^0}  \left[ \left|m^*_{\pi_n^0}F^*_{\pi_n^0}+N_{\pi_n^0}^*\right|^2 \right] &=- m_q\langle \bar qq\rangle^*, \\ 
    \frac{1}{2} \left( \sum_{n^+} \left[ \left|m^*_{\pi_n^+}F^*_{\pi_n^+}+N_{\pi_n^+}^*\right|^2 \right] + \sum_{n^-} \left[ \left|m^*_{\pi_n^-}F^*_{\pi_n^-}+N_{\pi_n^-}^*\right|^2 \right] \right)&=- m_q\langle \bar qq\rangle^*,
\end{align}
\end{subequations}
\end{widetext}
where $n^{\pm,0}$ label $\pi^{\pm,0}$ excited states. These are the generalized GOR relations in isospin asymmetric nuclear matter, which are detailed in \cref{app:gor}. Note that the $p_0$ dependence in the left-hand sides of these sum rules disappears when two correlation functions are summed up. In principle, all of the excited modes should contribute to the sum rules. From the argument in Ref.~\cite{Jido2008}, however, one can estimate the order of the masses and coupling constants in \cref{eq:gorre} in terms of the quark mass from the PCAC relation. Consider the PCAC relation given as \cref{eq:pcacim} for the excited states of the pions in the rest frame, where the masses are no longer small quantities. Then, the orders of the form factors are initiated from the $\mathcal{O}(m_q)$ order. Thus, one can expect that the contributions of the excited states begin at the $\mathcal{O}(m_q^2)$ order in \cref{eq:gorre}. Therefore, we use only the lowest states of the pions to estimate the reduction of the quark condensate in this paper. Note that $N_{1\pi}^*$ has the order of $\mathcal{O}(\sqrt{m_q})$ in this consideration. However, $N_{1\pi^{\pm}}^*$ are not zero in asymmetric nuclear matter in the chiral limit from the explicit calculation. To be consistent in the chiral limit, the symmetries and Nambu-Goldstone bosons in the chiral limit should be carefully considered, e.g., such as in Refs.~\cite{watanabe2012unified,hidaka2013counting}, which are beyond our current scope.

\section{Chiral perturbation theory in nuclear matter}\label{sec:methods}
We use in-medium chiral perturbation theory with a $SU(2)$ chiral Lagrangian to calculate the in-medium pion properties. In our scope, we consider the chiral Lagrangian up to second order in the chiral counting. We briefly discuss the in-medium chiral perturbation theory in \cref{sec:ChPT} and the detailed form of the chiral Lagrangian in \cref{sec:Lagrangian}. For a more thorough presentation, we refer to Refs.~\cite{Meisner2002,Goda2014,Huebsch2021}.

\subsection{In-medium chiral perturbation theory}\label{sec:ChPT}
Chiral perturbation theory (ChPT) is an effective field theory based on the symmetries of QCD, and formulated in terms of a chiral Lagrangian. 
In order to calculate in-medium quantities, we follow the in-medium ChPT formalism initiated by Oller~\cite{Oller2002}, and further developed by Ref.~\cite{Meisner2002}. 
The generating functional in quantum field theory is defined as the transition amplitude between two asymptotic ground states: $Z[J] = \langle \Omega_\text{out}|\Omega_\text{in}\rangle$.
We assume these asymptotic states to be described by Fermi seas of non-interacting nucleons:
\begin{align}
  |\Omega\rangle = \prod\limits_{i}^{\vec |p_i|\le k_F^{p,n}} a^\dagger (\vec p_i) |0\rangle,
\end{align}
where we include all momenta up to the Fermi momentum, which depends on the nucleon density as:
\begin{align}
    k_F^{p,n} = (3\pi^2 \rho_{p,n})^{1/3}. 
\end{align}
Furthermore, $a^\dagger (\vec p_i)$ is a nucleon creation operator with momentum $\vec p_i$, where $i$ denotes the isospin and spin of the nucleon, and $|0\rangle$ is the zero-particle vacuum state. 
Interactions between these nucleons, which can be important from $\mathcal O(\rho^2)$ in the density expansions, can be implemented by a nucleon--nucleon contact interaction term in the Lagrangian.

In the path integral formalism, the generating functional $Z[J]$ can be expressed using the Lagrangian of the system,
\begin{align}
  Z[J] &= \int \mathcal D U \mathcal D \bar N \mathcal D N \langle\Omega_\text{out}|N_{t\to+\infty}\rangle \nonumber\\
  &\qquad\times \eu^{\im \int \dd[4]x (\mathcal L_\pi + \mathcal L_N + \mathcal L_{\pi N})} \langle N_{t\to-\infty}|\Omega_\text{in}\rangle,
\end{align}
where proton and neutron fields are collected in the two-component isospin doublet $N=(p, n)^T$ and the full Lagrangian contains a pion term, a nucleon term, and a pion--nucleon interaction term. 

The nucleon field appears bilinear in the Lagrangian, which means it can be integrated out, as demonstrated by Ref.~\cite{Oller2002}.  
This results in an expansion of Fermi sea insertions, 
\begin{widetext}
\begin{align} \label{eq:Z}
	Z[J] = \int \mathcal DU \exp\bigg\{ \im\int\dd[4]x \Big[ \mathcal L_{\pi} &- \int\widetilde{\dd {p}}\, \mathcal{FT} \Tr{\im\Gamma(x,y)(\slashed p + m_N)n(\vec p)} \nonumber \\
	& -\frac{\im}{2} \int \widetilde{\dd {p}}\,\widetilde{\dd {q}}\, \mathcal{FT} \Tr{\im\Gamma(x,x')(\slashed q + m_N)n(\vec q) \im\Gamma(y',y) (\slashed p + m_N) n(\vec p)} + \ldots  \Big] \bigg\} ,
\end{align}
\end{widetext}
where $\widetilde{\dd p} = \dd[3]\vec p\, (2\pi)^{-3} (2E(\boldsymbol{p}))^{-1}$ denotes a Lorentz invariant integration measure, $E(\boldsymbol{p}) = (\vec p^2 + m_N^2)^{1/2}$ is the nucleon energy for momentum $\vec p$, $\mathcal{FT}$ represents a Fourier transformation of spatial variables (except for $x$) and $\Gamma(x,y)$ is a non-local vertex defined via in-vacuum quantities, 
\begin{align}\label{eq:gamma-definition}
	\im\Gamma &= A [ \id_4 - G_0 A ]^{-1}.
\end{align}
Here, the interaction operator $A$ and the free nucleon propagator $G_0$ are given by in-vacuum chiral perturbation theory. 
The matrix $n(\vec p)$ is used to implement the different Fermi momenta of protons and neutrons, 
\begin{align}
	n(\vec p) &= \begin{pmatrix}
		\Heavi(k_F^p - |\vec p|) & 0 \\
		0 & \Heavi(k_F^n - |\vec p|) 
	\end{pmatrix},
\end{align}
and restricts the momentum integration up to these Fermi momenta using Heaviside step functions. 
We will abbreviate these step functions as follows: $\Heavi_{\vec p}^{p,n} \equiv \Heavi(k_F^{p,n} - |\vec p|)$. 

In summary, the expression of the generating functional in \cref{eq:Z} represents three independent expansion schemes. 
First, there is an expansion in in-medium nucleon propagators, of which the first two terms are explicitly written in \cref{eq:Z}. 
Second, the non-local vertex $\Gamma$ can be expanded in terms of the pion-nucleon interaction operator $A$ as: $\im\Gamma = A + AG_0A + AG_0AG_0A + \ldots$. 
Finally, the pion-nucleon interaction operator $A$ is expanded as $A = A^{(1)} + A^{(2)} + \ldots$, where $A^{(i)}$ is counted as $\mathcal O(p^i)$ in the usual counting scheme of ChPT. 
In this work, we will include next-to-leading order effects from each of the three expansions, followed by removing any diagram that counts as $\mathcal O(k_F^6) = \mathcal O(\rho^2)$ in the density expansion. 
The reason for this exclusion is that at the order of $\mathcal O(\rho^2)$, nucleon-nucleon correlations become important, which are not present in this work. 

\subsection{The chiral Lagrangian}\label{sec:Lagrangian}
We employ a chiral Lagrangian up to second order in the chiral counting. 
There are two contributions to this Lagrangian: 
First, the pion Lagrangian, 
\begin{align}\label{eq:pionLag}
	\mathcal L_\pi ^{(2)} = \frac{f^2}{4} \Tr{D_\mu U^\dagger D^\mu U + \chi^\dagger U+\chi U^\dagger},
\end{align}
which describes pion-pion interactions, as well as interactions of pions with external currents via the covariant derivative and the field $\chi$. 
The chiral field $U$ is given in terms of the pion fields $\vec\pi=\pi^a \tau^a$:
\begin{align}
	U &= \exp(\im \vec\pi \frac{y(\pi^2)}{2\sqrt{\pi^2}})\qquad (\pi^2 = \pi^a\pi^a).
\end{align}
Note that the function $y(\pi^2)$ fulfills:
\begin{align}
	y(\pi^2)-\sin(y\left(\pi^2\right)) = \frac{4}{3} \left(\frac{\pi^2}{f^2}\right)^{3/2}.
\end{align}
This parametetrization of the chiral field enables a simple treatment of perturbative calculations~\cite{Charap1970,Charap1971,Gerstein1971}: 
the generating functional, \cref{eq:Z}, is defined in terms of the chiral field $U$, so when considering loop corrections to the pion field, one should be careful about chiral invariance of the integral measure. 
For a more thorough discussion of this Lagrangian, we again refer to Refs.~\cite{Goda2014,Huebsch2021}. 

The $\pi N$ interaction Lagrangian is given by $\mathcal L _{\pi N} = -\sum_{j=1}\bar N A^{(j)} N$, where $j$ denotes the chiral order. 
The terms which are relevant to the present calculation read:
\begin{subequations}
\begin{align}
    A^{(1)} &= -\im \gamma^\mu \Gamma_\mu - \im g_A \gamma^\mu \gamma^5 \Delta_\mu, \label{eq:A1} \\
    A^{(2)} &= -c_1 \Tr{\chi_+} + \frac{c_2}{2m_N^2}\Tr{u_\mu u_\nu} D^\mu D^\nu \nonumber\\
    &\qquad - \frac{c_3}{2}\Tr{u_\mu u^\mu} + \frac{c_4}{2} \gamma^\mu \gamma^\nu [u_\mu, u_\nu] + \ldots,  \label{eq:A2}
\end{align}
\end{subequations}
with
\begin{align}
    &D_\mu\psi=(\partial_\mu+\Gamma_\mu)\psi,\nonumber\\
    &\Gamma_\mu=\frac{1}{2}[u^\dagger,\partial_\mu u]-\frac{i}{2}u^\dagger(v_\mu+a_\mu)u-\frac{i}{2}u(v_\mu-a_\mu)u^\dagger,\nonumber\\
    &\Delta_\mu=\frac{1}{2}\{u^\dagger[\partial_\mu-i(v_\mu+a_\mu)]u-u[\partial_\mu-i(v_\mu-a_\mu)]u^\dagger\},\nonumber\\
    &u_\mu=2i\Delta_\mu,\nonumber\\
    &\chi_\pm=u^\dagger\chi u^\dagger\pm u\chi^\dagger u.\nonumber
\end{align}
The detailed forms of interactions arising from this Lagrangian are listed in \cref{app:interactions,app:interactions_physical}.

\begin{table} \centering
\begin{tabular}{@{}llllll@{}}\toprule
	& LEC & Set 1\cite{Scherer2012} & Set 2\cite{Goda2014} & Set 3\cite{Aoki2019PhD} \\\midrule
	& $c_1$ & $-0.90$ & $-0.59$ & $-0.61$ \\
	& $c_2$ & $\ \ 3.30$  & $\ \ 3.30$ &  $\ \ 2.97$ \\
	& $c_3$ & $-4.70$ & $-4.43$ & $-4.05$ \\\bottomrule
\end{tabular}
\caption{The $SU(2)$ low-energy constants (LECs) $c_1, c_2$ and $c_3$ in  units of $10^{-3}$ MeV$^{-1}$. The set 1,2, and 3 are taken from Refs.~\cite{Scherer2012,Goda2014,Aoki2019PhD}, respectively.}
\label{table1}
\end{table}

In this work, we assume that in-vacuum effects are renormalized by counter-terms in the chiral Lagaragian, so we use physical values for the in-vacuum quantities. We put on physical values for the in-vacuum pion decay constant $f_\pi = 92.4$ MeV, and the axial coupling $g_A = 1.26$. 
We also use isospin-averaged values for the in-vacuum pion mass, $m_\pi = 138$ MeV, and the nucleon mass, $m_N=939$ MeV. 
For the low-energy constants (LECs) that appear in the pion-nucleon interaction Lagrangian, we use sets 1, 2, and 3 in \cref{table1} sourced from Refs.~\cite{Scherer2012,Goda2014, Aoki2019PhD}. In this paper, the numerical value of $c_4$ is not needed because corrections proportional to $c_4$ are higher-order contributions beyond the scope of our study. 

\section{Results}\label{sec:results}
In this section, we show the Feynman diagrams that we used to compute the in-medium pion properties. In \cref{sec:self-energy}, we calculate the pion self-energies, which we use to compute the in-medium pion masses in \cref{sec:mass}, and the wave function renormalizations in \cref{sec:wfr}. 
In \cref{sec:decay-constant}, we use another set of diagrams to compute the vertex corrections for the in-medium pion decay constants, which together with the wave function renormalizations yield the in-medium decay constants. 
In \cref{sec:GOR}, we derive an in-medium Gell-Mann--Oakes--Renner relation, and investigate whether our results fulfill this equation. Up to this point in the calculation of the in-medium pion properties, we use the physical value of the in-vacuum pion mass and the sets of LECs in \cref{table1}. 

\begin{figure*}%
\subfloat[]{{\includegraphics[width=0.4\textwidth ]{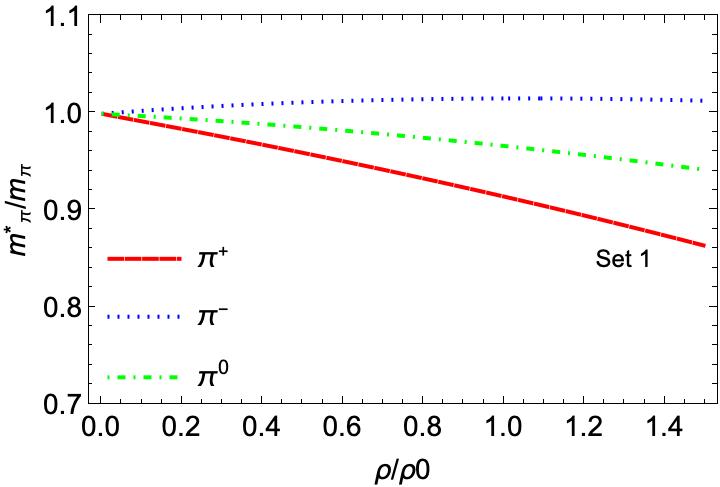} }}%
\subfloat[]{{\includegraphics[width=0.4\textwidth ]{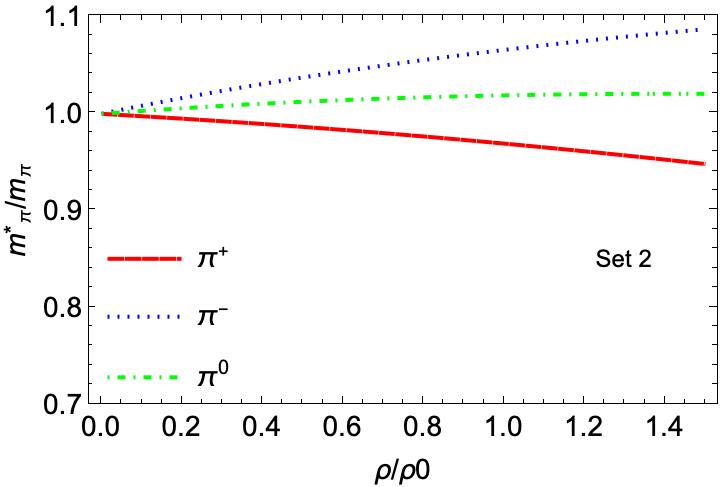} }}%
\hfill 
\subfloat[]{{\includegraphics[width=0.4\textwidth ]{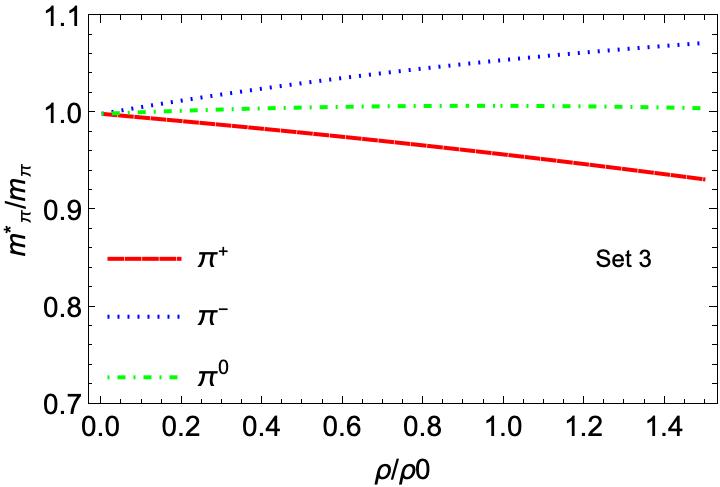} }}%
\subfloat[]{{\includegraphics[width=0.4\textwidth ]{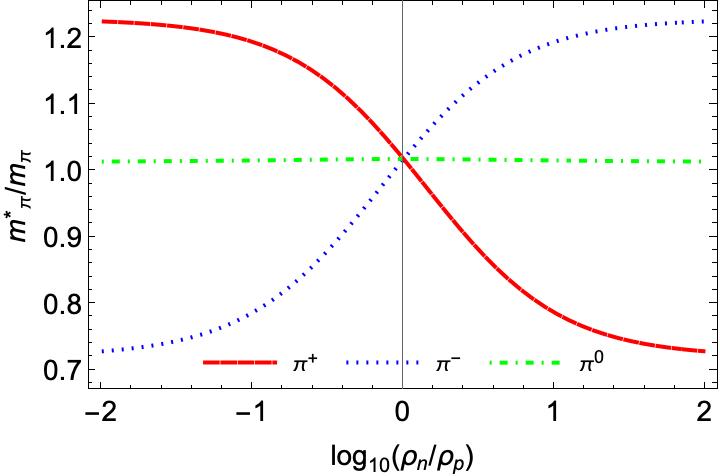} }}%
\caption{The triplet pion masses split under the influence of isospin-asymmetric nuclear matter. From (a) to (c), they are referred to as sets 1-3, respectively at $r=1.5$. (d) shows the ratio of $\rho_n/\rho_p$ dependence of the pion masses in set 2 of the LECs at normal nuclear density $\rho=\rho_0$. All of them are in units of the isospin-averaged vacuum pion mass.}%
\label{fig1}%
\end{figure*}

\subsection{Self-energies}\label{sec:self-energy}
In order to compute the pion self-energies, we use Feynman diagrams with two external lines corresponding to pion propagators (drawn as dashed lines) having momentum $q^\mu$. 
According to the in-medium expansion discussed in \cref{sec:ChPT}, we consider all diagrams with up to two in-medium nucleon propagators (denoted by a thick line), up to one free nucleon propagator (denoted by a thin line) between two in-medium vertices, and use pion-nucleon interactions from $A^{(1)}$ and $A^{(2)}$. 
In total, we consider these ten diagrams: 
\begin{align}
	-\im\Sigma_{\pi}(q) &=\resizebox{\dThree}{!}{$\diagram{
		\SelfEnergyFirstA;
	}$} + \resizebox{\dThree}{!}{$\diagram{
		\SelfEnergyFirstB;
	}$}\nonumber\\
    &+ \resizebox{\dTwo}{!}{$\diagram{
		\SelfEnergyTypeTwo;
		\vertex{B}{1};
	}$} + \resizebox{\dTwo}{!}{$\diagram{
		\SelfEnergyTypeTwo;
		\vertex{B}{2};
	}$}\nonumber\\
    &+ \resizebox{\dTwo}{!}{$\diagram{
		\SelfEnergyTypeThree;
		\vertex{B}{2};
		\vertex{E}{1};
		\vertex{F}{1};}
	$}
        + \resizebox{\dThree}{!}{$\diagram{
		\SelfEnergyTypeFour;
		\vertex{B}{1};
            \vertex{C}{1}
	}$}\nonumber\\
    &+\resizebox{\dThree}{!}{$\diagram{
		\SelfEnergyTypeFour;
		\vertex{B}{1};
            \vertex{C}{2}
	}$} +\resizebox{\dThree}{!}{$\diagram{
		\SelfEnergyTypeFour;
		\vertex{B}{2};
            \vertex{C}{1}
	}$}\nonumber\\
    &+\resizebox{\dThree}{!}{$\diagram{
		\SelfEnergyTypeFour;
		\vertex{B}{2};
            \vertex{C}{2}
	}$} + \resizebox{\dTwo}{!}{$\diagram{
		\SelfEnergyTypeFive;
		\vertex{B}{1};
		\vertex{D}{1};
	}$}.
    \label{self-energy diagrams}
\end{align}
Note that the first and second diagrams together correspond to a diagram with Pauli-blocked nucleon propagators, where the nucleon momentum (either $p^\mu$ or $k^\mu$) must be above the Fermi momentum $k_F$. For more details, we refer to Ref.~\cite{Goda2013}, where this correspondence was explicitly shown. 

After applying the Feynman rules for in-medium ChPT~\cite{Meisner2002}, we can obtain the integral form for each diagram, which is shown in \cref{app:Selfenergy_Decayconst}.

We will show our results in terms of $\rho/\rho_0$, which is a convenient normalization of the nucleon density to the normal nuclear density $\rho_0 \approx 0.17$ fm${}^{-3}$. 

\subsection{Masses}\label{sec:mass}

We calculate the in-medium pion mass by solving \cref{eq:pionmass}. Up to the order of the density we are considering in this study, we can evaluate the self-energy by substituting $q_0 \to m_\pi$~\cite{Goda2014}. We call this procedure a perturbative calculation. This allows us to write simply the in-medium pion mass as
\begin{align}
    m^*_\pi = \sqrt{m_\pi^2 + \Sigma_\pi (q_0=m_\pi, \vec 0)}.
\end{align}
The following results are calculated using this procedure. At the end of this section, we will discuss the difference between the perturbative calculation and the results obtained by using the solutions of the mass equation \cref{eq:pionmass} in the leading order to the self-energies. We call the results from the latter procedure partially perturbative results of the in-medium masses.

We show the density dependence of the in-medium pion masses and their dependence on the nucleon ratio in \cref{fig1}(a)--(d). 
\cref{fig1}(a)--(c) show the three pion masses and their dependence on the nucleon density at a fixed ratio $r=\rho_n/\rho_p=1.5$ for the LECs sets 1,2 and 3. The mass of the negatively charged pion increases at normal nuclear density and for a nucleon ratio of $r=1.5$ by 2--7\%, depending on the choice of the LECs. 
This is in quite good agreement with other works that computed the same in-medium mass: 
Refs.~\cite{Drews2015,Drews2017} use functional renormalization group methods and report an increase of 6\%. 
Ref.~\cite{Kaiser2001} uses two-loop chiral perturbation theory and arrives at an increase of 10\%. 
Finally, Ref.~\cite{Meisner2002} uses in-medium chiral perturbation theory up to next-to-leading order and reports an increase of 13\%. 
The mass of the neutral pion decreases by 3\%, and increases by 2\% and 1\% for sets 1,2 and 3, respectively.
We also found decreases by 4--8\% of the positive pion mass according to the sets of LECs.
We can compare these values to Ref.~\cite{Kaiser2001} which reports an increase of 4\% and a decrease of 1\%, respectively. In summary, for neutron-rich nuclear matter, the $\pi^-$ mass is increased while the $\pi^+$ mass is decreased. 
For proton-rich nuclear matter, the opposite is the case.

\begin{table*}\centering
\begin{tabular*}{0.99\textwidth}{@{}llllll@{}}\toprule
	& & $m_{\pi^-}^*/m_\pi$ & $m_{\pi^0}^*/m_\pi$ & $m_{\pi^+}^*/m_\pi$ & $m_{\pi}^*/m_\pi (r=1)$\\\midrule
	& This work (perturbative) &  & &  &  \\
        & Set 1 & $1.02$ & $0.97$ & $0.92$ & $0.97$ \\
        & Set 2 & $1.07$ & $1.02$ & $0.97$ & $1.02$ \\
        & Set 3 & $1.06$ & $1.01$ & $0.96$ & $1.01$ \\
        & This work (partially perturbative) &  & &  &
    \\
        & Set 1 & $1.05$ & $0.94$ & $0.84$ & $0.94$ \\
        & Set 2 & $1.13$ & $1.04$ & $0.95$ & $1.04$ \\
        & Set 3 & $1.10$ & $1.02$ & $0.93$ & $1.02$ \\
	Th. & Ref.~\cite{Kaiser2001} & $1.10$ & $1.04$ & $0.99$ & --- \\
	& Refs.~\cite{Drews2015,Drews2017} & $1.06$ & --- & --- & --- \\
	& Ref.~\cite{Meisner2002} & $1.13$ & --- & --- & --- \\
	& Ref.~\cite{Hutauruk2019} & --- & --- & --- & $0.94$\\
	& Ref.~\cite{Wirzba1995} & --- & --- & --- & $1.03$\\
	Ex. & Ref.~\cite{Itahashi2000} & $1.16\sim 1.19$ & --- & --- & ---
     \\\bottomrule
\end{tabular*}
\caption{We compare our results of the in-medium pion masses at $r=1.5$ (first three values) and at $r=1$ (last column) to theoretical (Th.) and experimental (Ex.) results. All values correspond to densities at normal nuclear density: $\rho=\rho_0$ and are given in units of the isospin-averaged vacuum pion mass.}
\label{tab:res:m}
\end{table*}

For isospin-symmetric nuclear matter, the masses of all pions decrease equally by 3\% and increase by 1--2\% for each set of LECs at normal nuclear density. This is compatible with the results in Ref.~\cite{Hutauruk2019}, which uses an NJL model and reports a decrease of 5\%, and  Ref.~\cite{Wirzba1995}, which uses the chiral perturbation theory and reports an increase of 3\%.

As one of the main objectives of this work is to investigate the effect of isospin-asymmetric nuclear matter, we plot the density dependence of the in-medium pion masses over a range of neutron-to-proton ratios in \cref{fig1}(d), which shows the three pion masses and how they depend on the ratio $r$ at fixed nuclear density of $\rho=\rho_0$.
The far right (left) of \cref{fig1}(d) corresponds to a neutron (proton) surplus, and we observe the behaviors of $m_{\pi^{\pm,0}}^*(r)$ are symmetric around $r=1$, such that the $\pi^0$ mass is invariant under the exchange $r\leftrightarrow 1/r$, while $\pi^\pm \to \pi^\mp$. As shown in \cref{fig1}(d), when we consider a symmetric nuclear matter i.e., a neutron-to-proton ratio of 1, the three pion masses are equal.

We represent our results in~\cref{tab:res:m}. Note that the partially perturbative results are closer to those of other references. This is because higher-order contributions are included in these results. 


\subsection{Wave function renormalizations}\label{sec:wfr}
The wave function renormalizations are calculated using \cref{eq:def-Z} with the self-energies in the rest frame of the pions. Note that we get some divergence terms in the wave function renormalizations from two-loop contributions in the limit $q_0\rightarrow m_\pi$, which must be canceled in physical $S$-matrix elements. Thus, we drop these divergence terms. The detailed argument is given in \cref{ap:singular}.

\begin{figure*}%
\subfloat[]{{\includegraphics[width=0.4\textwidth ]{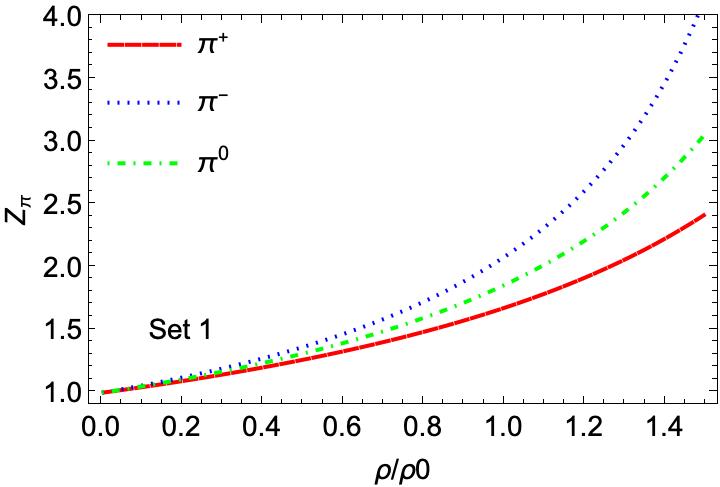} }}%
\subfloat[]{{\includegraphics[width=0.4\textwidth ]{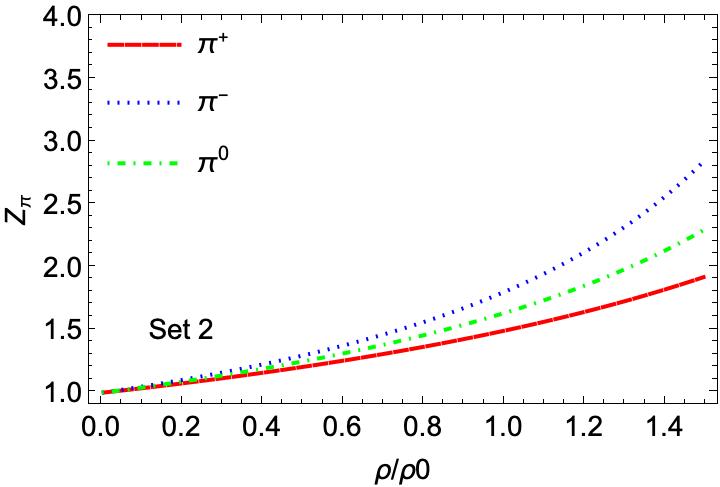} }}%
\hfill 
\subfloat[]{{\includegraphics[width=0.4\textwidth ]{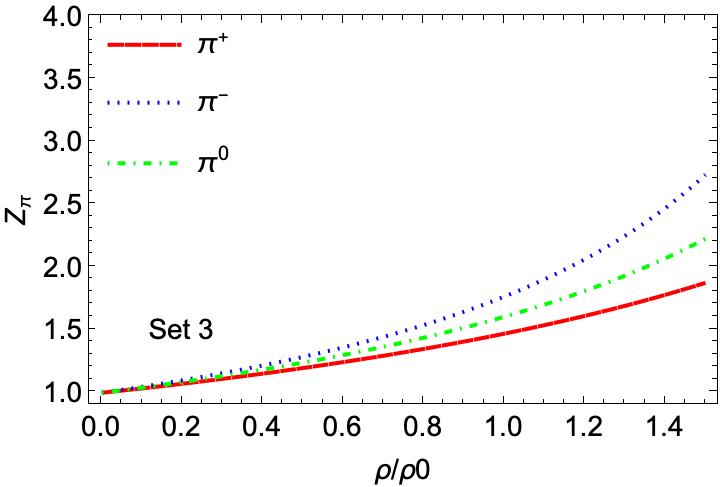} }}%
\subfloat[]{{\includegraphics[width=0.4\textwidth ]{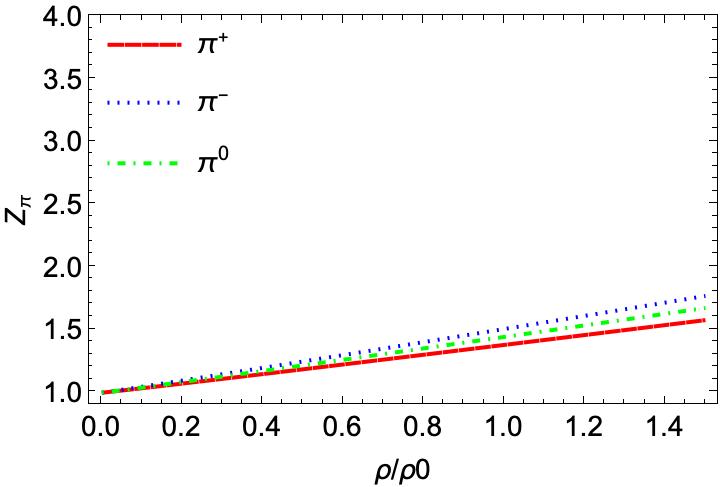} }}%
\caption{The behaviors of the wave function renormalizations of the three pions according to the sets of LECs are shown in (a)--(c). (d) shows the behavior of the wave function renormalizations using the small $k_F$ expansion in set 2 of the LECs. All plots are evaluated at $r=1.5$}%
\label{fig2}%
\end{figure*}

The density dependences of the in-medium pion wave function renormalizations are shown in \cref{fig2}(a)--(c) for the isospin-asymmetric nuclear matter of $r=1.5$. \cref{fig2}(d) shows the behavior of the wave function renormalizations using the small $k_F$ expansion with set 2 of the LECs at $r=1.5$. All three pion wave function renormalizations increase in nuclear matter. At normal nuclear density, the in-medium wave function renormalization increases by 77--101\% for the negatively charged pion, 61--85\% for the neutral pion, and 47--68\% for the positively charged pion.

In the case of isospin-symmetric nuclear matter, the wave function renormalization of all pions increases by around 61--86\% at normal nuclear density. We can compare this to other works, for instance Ref.~\cite{Chanfray2003}, which used the chiral perturbation theory approach and connected those results to a linear sigma model calculation, and reported an increase of 51--53\%. Furthermore, Ref.~\cite{Goda2014} reports a 40\% increase at the normal nuclear density. We summarize our results for the wave function renormalization in \cref{tab:res:z}.



\begin{table}\centering
\begin{tabular*}{0.46\textwidth}{@{}llllll@{}}\toprule
	& & $Z_{\pi^-}$ & $Z_{\pi^0}$ & $Z_{\pi^+}$ & $Z_{\pi} (r=1)$\\\midrule
	& This work &  &  &  &  \\
        & Set 1 & $2.1$ & $1.85$ & $1.68$ & $1.86$ \\
        & Set 2 & $1.81$ & $1.63$ & $1.49$ & $1.63$ \\
        & Set 3 & $1.77$ & $1.61$ & $1.47$ & $1.67$ \\
	Th. & Ref.~\cite{Chanfray2003} & --- & --- & --- & $1.51\sim1.53$ \\
	& Ref.~\cite{Goda2014} & --- & --- & --- & $1.40$
    \\\bottomrule
\end{tabular*}
\caption{We compare our results of the wave function renormalization at $r=1.5$ (first three values) and at $r=1$ (last column) to theoretical (Th.) results. All values correspond to normal nuclear density.}
\label{tab:res:z}
\end{table}

\subsection{Decay constants}\label{sec:decay-constant}
We compute the effective vertex corrections defined in \cref{eq:effvex} using Feynman diagrams with one incoming pion line and one external axial-vector current (denoted by a wavy line). Then, the in-medium temporal pion decay constants can be calculated using \cref{eq:fd} with wave function renormalizations. 
In total, we consider these ten diagrams:
\begin{align}
	\im q_0 \hat f_{\pi} &= 
	\resizebox{\dTwo}{!}{$\diagram{
		\DecayTypeTwo;
		\vertex{B}{1};
	}$} + \resizebox{\dTwo}{!}{$\diagram{
		\DecayTypeTwo;
		\vertex{B}{2};
	}$}\nonumber\\
    &+ \resizebox{\dThree}{!}{$\diagram{
		\DecayConstantFirstA;
	}$} + \resizebox{\dThree}{!}{$\diagram{
		\DecayConstantFirstB;
	}$}\nonumber\\
    &+ \resizebox{\dTwo}{!}{$\diagram{
		\DecayTypeThree;
		\vertex{B}{2};
		\vertex{E}{1};
		\vertex{F}{1};}
	$} + \resizebox{\dThree}{!}{$\diagram{
		\DecayTypeFour;
		\vertex{B}{1};
		\vertex{C}{1};
	}$}\nonumber\\
    &+ \resizebox{\dThree}{!}{$\diagram{
		\DecayTypeFour;
		\vertex{B}{1};
		\vertex{C}{2};
	}$} + \resizebox{\dThree}{!}{$\diagram{
		\DecayTypeFour;
		\vertex{B}{2};
		\vertex{C}{1};
	}$}\nonumber\\
    &+ \resizebox{\dThree}{!}{$\diagram{
		\DecayTypeFour;
		\vertex{B}{2};
		\vertex{C}{2};
	}$} + \resizebox{\dTwo}{!}{$\diagram{
		\DecayTypeFive;
		\vertex{B}{1};
		\vertex{D}{1};
	}$}. 
    \label{axial 1PI diagrams}
\end{align}
We can obtain integral forms of these diagrams using the Feynman rules similar to the self-energies. We show the explicit expressions of the diagrams in \cref{app:Selfenergy_Decayconst}.
\begin{figure*}%
\subfloat[]{{\includegraphics[width=0.4\textwidth ]{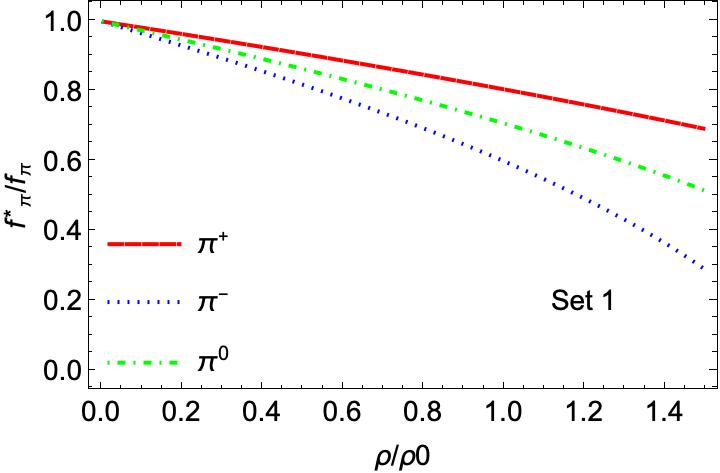} }}%
\subfloat[]{{\includegraphics[width=0.4\textwidth ]{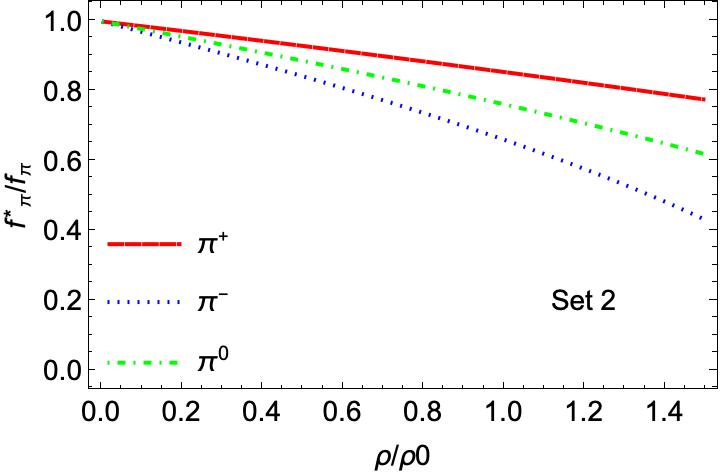} }}%
\hfill 
\subfloat[]{{\includegraphics[width=0.4\textwidth ]{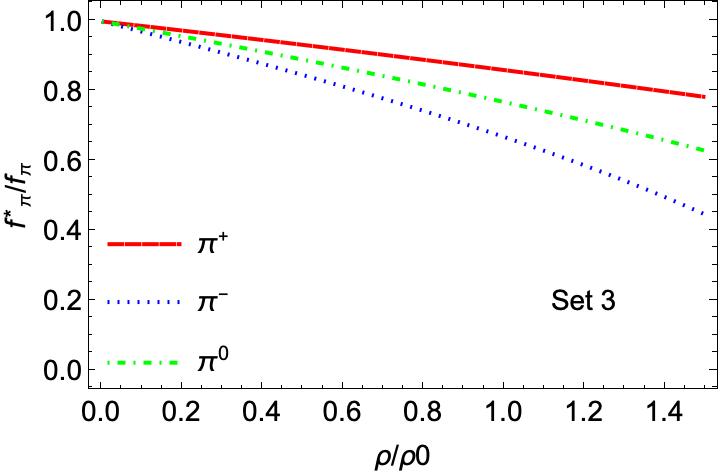} }}%
\subfloat[]{{\includegraphics[width=0.4\textwidth ]{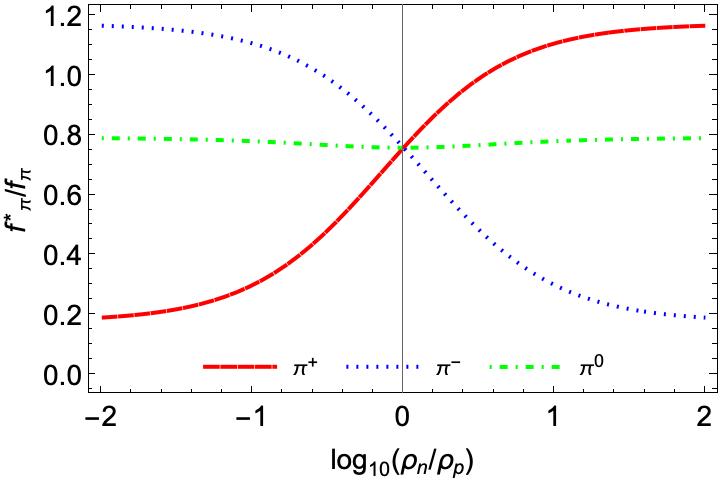} }}%
\caption{The in-medium pion decay constants of the three pions. The ratios of in-medium to vacuum pion decay constants at the ratio of $\rho_n/\rho_p=1.5$ depending on the nucleon density for the sets 1--3 are plotted at (a)--(c) . (d) shows the ratio of $\rho_n/\rho_p$ dependence of the in-medium pion decay constants in set 2 of the LECs at the normal nuclear density $\rho=\rho_0$. As with the case of the masses, the behaviors of $f_{\pi^{\pm,0}}^*(r)$ are symmetric around $\rho_n/\rho_p=1$, such that $f_{\pi^0}^*$ is symmetric under the exchange $r\leftrightarrow 1/r$, while $f_{\pi^\pm}^* \to f_{\pi^\mp}^*$.}%
\label{fig3}%
\end{figure*}

\begin{table}\centering
\begin{tabular*}{0.46\textwidth}{@{}llllll@{}}\toprule
	& & $f_{\pi^-}^*/f_{\pi}$ & $f_{\pi^0}^*/f_{\pi}$ & $f_{\pi^+}^*/f_{\pi}$ & $f_{\pi}^*/f_{\pi} (r=1)$\\\midrule
	& This work &  &  &  &  \\
        & Set 1 & $0.79$ & $0.83$ & $0.89$ & $0.83$ \\
        & Set 2 & $0.82$ & $0.86$ & $0.91$ & $0.86$ \\
        & Set 3 & $0.82$ & $0.87$ & $0.91$ & $0.87$ \\
	Th. & Ref.~\cite{Jido2008} & --- & --- & --- & $0.89$ \\
	Ex. & Ref.~\cite{Kienle2004} & $0.85\sim0.91$ & --- & --- & --- \\ 
        & Ref.~\cite{nishi2023chiral} & $0.84\sim0.89$ & --- & --- & ---
    \\\bottomrule
\end{tabular*}
\caption{We compare our results of the in-medium pion decay constant at $r=1.4$ (first three values) and at $r=1$ (last column) to theoretical (Th.) and experimental (Ex.) results. All values correspond to densities at the density of $\rho\approx0.6\rho_0$ and are given in units of the isospin-averaged vacuum pion decay constant. }
\label{tab:res:f}
\end{table}

For accessibility of experiments, we compare our results for decay constants at the density $\rho=0.6\rho_0$ and the ratio of $\rho_n/\rho_p=1.4$. We get decreases in the decay constant of 
the negatively charged pion by 18--21\% according to the choice of LECs. In Ref.~\cite{nishi2023chiral}, a decrease of the negatively charged pion decay constant is reported by around 11--16\%, where the spectrum of pionic Sn atoms was measured. Reference.~\cite{Kienle2004} shows that the decay constant of the negative pion decreases around by 9--15\%.
Furthermore, our calculations show that the decay constants for the positive pion decrease by around 9--11\% and for the neutral pion by around 13--17\%, depending on the choice of the LECs. 

For isospin-symmetric nuclear matter, all three decay constants behave the same and decrease by around 13--17\% at the density of $\rho\approx0.6\rho_0$. In Ref.~\cite{Jido2008}, the in-medium pion decay constant was reported to decrease by 11\% under the same nuclear matter conditions, and the parameter $b_1^*$ was able to be determined, which is used to parametrize the isovector part of the $s$-wave pion-nucleus optical potential. 
The numerical value of $b_1^*$ was obtained from the pionic atom and $\pi^-$-nucleus scattering data. 
The following relation between $b_1^{(*)}$ and $f^{(*)}_\pi$: $b_1/b_1^* = (f^*_\pi/f_\pi)^2$ was found, and since the in-medium quantity $b_1^*$ was found to be enhanced in nuclear matter, it was concluded that the in-medium pion decay constant would be reduced in nuclear matter. 
We also compare our results for the isospin-symmetric case further at normal nuclear density $\rho=\rho_0$ with other theoretical references. In this case, the three pion decay constants are reduced by 23--29\%. 
This agrees well with other works (in the following all values are given at $\rho=\rho_0$), e.g. Ref.~\cite{Goda2014}, which reported a 25\% decrease and Refs.~\cite{Meisner2002,Lacour2010}, which used in-medium chiral perturbation theory up to next-to-leading order to arrive at a 26\% decrease.
Furthermore, Ref.~\cite{Hutauruk2019} used an NJL model and reports a decrease of 13\% and Ref.~\cite{Kienle2004} reports a decrease of 20\% by deriving $s$-wave pion nucleus potential parameters.

The results of the decay constants evaluated at $q_0=m_\pi$ are shown in~\cref{tab:res:f}. Note that the results using the partially perturbative results of the in-medium masses are within 1\% error from the perturbative results. We also represent the behaviors of the decay constants according to densities with $r=1.5$ for the sets 1,2 and 3 of the LECs in \cref{fig3}(a)--(c). The behavior in the ratios of $\rho_n/\rho_p$ at $\rho=\rho_0$ is displayed in \cref{fig3}(d) with set 2 of the LECs.



\subsection{Gell-Mann--Oakes--Renner relations}\label{sec:GOR}
Using the results from \cref{sec:mass,sec:decay-constant}, 
we can investigate the accuracy of the in-medium GOR relations in the asymmetric nuclear matter within our formulation. For simplicity, we introduce the abbreviations: 
\begin{subequations}\label{eq:C_defs}
\begin{align}
    C_{ch} &\equiv \frac{|f_{\pi^+}^*|^2 m_{\pi^+}^{*2} + |f_{\pi^-}^*|^2 m_{\pi^-}^{*2}}{2},\\
    C_{ne} &\equiv |f_{\pi^0}^*|^2 m_{\pi^0}^{*2},\\ 
    C_{qq} &\equiv -    m_q\langle\bar qq\rangle_0,
\end{align}
\end{subequations}
where $\langle\bar qq\rangle_0$ is the quark condensate in vacuum, and we get the following results for set 3 of LECs at $\rho
=0.6\rho_0$ and $r=1.4$:
\begin{subequations}\label{eq:GOR-numerical-result}
\begin{align}
    C_{ch}/C_{ne} &= 1.00,\label{eq:GOR-numerical-result1}\\
    C_{ch}/C_{qq} &= 0.77,\label{eq:GOR-numerical-result2}\\
    C_{ne}/C_{qq} &= 0.77.
    \label{eq:GOR-numerical-result3}
\end{align}
\end{subequations}
Here we used the following values: $m_\pi = 138$ MeV, $f_\pi=92.4$ MeV, and $-m_q\langle\bar qq\rangle_0= m_\pi^2f_\pi^2$. We take only the lowest states of the pions for the GOR relations in \cref{eq:GOR-numerical-result}.
\cref{eq:GOR-numerical-result2,eq:GOR-numerical-result3} are in good agreement with Ref.~\cite{nishi2023chiral}, which showed 21--25\% reduction of the quark condensate in the nuclear medium.

\section{Summary}\label{sec:summary}
We have defined the in-medium pion properties using the correlation function approach in the physical basis of the pions. The in-medium pion mass is calculated by solving the self-consistent mass equation, and the coupling constants are obtained as residues at the pion pole positions of the correlation functions, which can be calculated by vertex corrections with the wave function renormalization, as pointed out in Refs.~\cite{Jido2008,hatsuda1999precursor}, in field theoretical approaches. In isospin-asymmetric nuclear matter, the positive and negative energy solutions and pole positions of the charged pions are different. Thus, the behaviors of the pion properties depend both on the density and the ratio $r=\rho_n/\rho_p$ of the nuclear matter.

The in-medium chiral perturbation theory has been used to compute in-medium pion properties with different Fermi momenta of the proton and neutron.
We have calculated Feynman diagrams for the in-medium self-energies, the 1PI vertex corrections to the pion decay constants. 
This enabled us to compute the in-medium masses, wave function renormalizations, and decay constants of the triplet pions. In neutron-dominant nuclear matter, the negative pion mass increases, but the positive pion mass decreases in the range of $\rho<1.5\rho_0$. In proton-dominant nuclear matter, the behaviors are reversed. On the other hand, the behavior of the neutral pion mass is symmetric under $r\leftrightarrow1/r$.

All the in-medium temporal pion decay constants decrease at $r$ regions close to isospin-symmetric nuclear matter such as $r=1.5$. In neutron-dominant nuclear matter at these ratios, the decay constant of the positive pion decreases slower than that of the negative pion. Again, the behaviors of the two pions are flipped under $r\leftrightarrow1/r$, and that of the neutral pion is symmetric around $r=1$.

The overall behaviors of the masses and the decay constants are symmetric around $r=1$, where all the results of the three pions are degenerate.

Furthermore, we have derived the generalized in-medium Gell-Mann–Oakes–Renner relations with all excited states of the pions, which are valid for isospin-asymmetric nuclear matter. The in-medium GOR relations can be approximated to the contributions of the lowest states of the pions under the $\mathcal{O}(m_q^2)$ order. 

Future works might include nucleon-nucleon interaction terms in the Lagrangian, and consider scatterings of a pion and an external nucleon. 
These would lead to new effects beyond the linear density and further discussion about singular parts in the wave function renormalizations. Also, the chiral limit case can be considered in future work, taking into account symmetries.

\section*{Acknowledgements}
The work of Y.\,S.~was supported by JST SPRING, Japan Grant Number JPMJSP2106 and JPMJSP2180. The work of D.\,J.~was partly supported by Grants-in-Aid for Scientific Research from JSPS (17K05449, 21K03530, 25K07315).

\appendix

\section{Detail calculations of the correlation functions  }\label{app:MediumGORPhysBasis}
In this appendix, we show the details of calculations of the correlation functions used in \cref{sec:in-medium-pion-properties,sec:in-medium-GOR}. We first consider the correlation function $\Pi^{ab}(q)$ in the physical basis for $a=\pm,b=\mp$ given as 
\begin{align}
    \Pi^{\pm\mp}(q) =  \int \dd[4]x \eu^{\im q \cdot x} \langle \Omega | \mathsf T P^\pm(x) P^\mp(0) | \Omega \rangle.
\end{align}
The correlation function in the coordinate space reads:
\begin{widetext}
\begin{align} \label{eq:apC1}
    \MoveEqLeft \langle\Omega|\mathsf T P^\pm (x) P^\mp(0)|\Omega\rangle = \nonumber\\
	&\sum\limits_n\int \frac{\dd[3]\vec p_n}{(2\pi)^32\omega_n} \bigg[ \Heavi(x_0) \langle\Omega|P^\pm (x) |n(p_n)\rangle\langle n(p_n)|P^\mp(0)|\Omega\rangle + \Heavi(-x_0) \langle\Omega|P^\mp(0) |n(p_n)\rangle\langle n(p_n)|P^\pm (x)|\Omega\rangle \bigg],
\end{align}
\end{widetext}
where we have inserted a complete set of one particle states. The inclusion of multiple particle states can be done straightforwardly. We use the matrix elements of the pseudoscalar current between one in-medium pion state and an asymptotic state in nuclear matter defined in~\cref{eq:P_coupling_medium,eq:P_coupling_medium_cc}.
Substituting them into each term in \cref{eq:apC1}, we obtain:
\begin{widetext}
\begin{align}
    \MoveEqLeft \langle\Omega|\mathsf T P^\pm (x) P^\mp(0)|\Omega\rangle =
	 \sum\limits_{n^\pm}\int \frac{\dd[3]\vec p_{n^\pm}}{(2\pi)^3} \Heavi(x_0) \frac{G^{*}_{\pi^{\pm}_n} \bar G^{*}_{\pi^\pm_n}(p_{n^\pm}) \eu^{- \im p_{n^\pm} \cdot x}}{2\omega_{n^\pm}} + \sum\limits_{n^\mp}\int \frac{\dd[3]\vec p_{n^\mp}}{(2\pi)^3} \Heavi (-x_0) \frac{\bar G^{*}_{\pi^\mp_n} G^*_{\pi^\mp_n}(p_{n^\mp}) \eu^{\im p_{n^\mp} \cdot x}}{{2\omega_{n^\mp}}} ,
\end{align}
where $n$ labels $\pi^\pm$ excited states, and $p^\mu_{n^{\pm}}=(\omega_{n^\pm}(\vec p_{n^{\pm}}),\vec p_{n^{\pm}})$. We can now use the following mathematical identity: 
\begin{align}
	\Heavi(\pm x_0)\eu^{\mp \im \omega_n x_0} &= \pm\frac{1}{2\pi\im} \int\limits_{-\infty}^{\infty}\dd{\omega}\frac{\eu^{\im\omega x_0}}{\omega \pm \omega_n \mp \im\epsilon}.
    \label{eq:mathematical_identity}
\end{align}
By using this, we write 
the correlation function $\Pi^{\pm\mp} (q)$ as 
\begin{align}
    \Pi^{\pm\mp} (q) &= -\im \int \dd[4]x \eu^{\im q\cdot x} \bigg[ \sum \limits_{n^\pm} \int \frac{\dd[3]\vec p_{n^\pm}\dd s}{
    (2\pi)^4} \frac{G^{*}_{\pi^{\pm}_n} \bar G^{*}_{\pi^\pm_n}(p_{n^\pm}) \eu^{\im (sx_0 + \vec p_{n^\pm} \cdot \vec x)}}{2\omega_{n^\pm}(s + \omega_{n^\pm} - \im \epsilon)} - \sum \limits_{n^\mp} \int \frac{\dd[3]\vec p_{n^\mp} \dd s}{
    (2\pi)^4} \frac{\bar G^{*}_{\pi^{\mp}_n} G^{*}_{\pi^\mp_n}(p_{n^\mp}) \eu^{\im (sx_0 - \vec p_{n^\mp} \cdot \vec x)}}{2\omega_{n\mp}(s - \omega_{n^\mp} + \im \epsilon)} \bigg].
\end{align} 
We perform the integration over $\dd[4] x$:
\begin{align}
    \Pi^{\pm\mp} (q) &= - \im  
    \bigg[ \sum \limits_{n^\pm} \int \dd[3] \vec p_{n^\pm} \dd s\frac{G^{*}_{\pi^{\pm}_n} \bar G^{*}_{\pi^\pm_n}(p_{n^\pm}) \delta^{(3)}(\vec p_{n^\pm} - \vec q)}{2\omega_{n^\pm}(s + \omega_{n^\pm} - \im \epsilon)} - \sum \limits_{n^\mp} \int \dd[3] \vec p_{n^\mp} \dd s \frac{\bar G^{*}_{\pi^{\mp}_n} G^{*}_{\pi^\mp_n}(p_{n^\mp}) \delta^{(3)}(\vec p_{n^\mp} + \vec q)}{2\omega_{n^\mp}(s - \omega_{n^\mp} + \im \epsilon)} \bigg]  \delta(q^0+s).
\end{align}
Also, the integration over $\dd[3] \vec p_n \dd s$ is now trivial and we obtain
\begin{align}
    \Pi^{\pm\mp} (q) &= \im \sum_{n^\pm} \frac{G^{*}_{\pi^\pm_n}(\vec q) \bar G^*_{\pi^\pm_n}(\vec q)}{2\omega_{n^\pm}(\vec q)(q^0 -\omega_{n^\pm}(\vec q) + \im \epsilon)} - \im \sum_{n^\mp} \frac{\bar G^{*}_{\pi^\mp_n}(-\vec q) G^*_{\pi^\mp_n}(-\vec q)}{2\omega_{n^\mp}(-\vec q)(q^0 + \omega_{n^\mp}(-\vec q) - \im \epsilon)},
\end{align}
where $\omega_{n^{\pm}}$ are the in-medium energy dispersion of the pions, and $\omega_\pi(\vec q)=\omega_\pi(-\vec q)$ and $G_\pi(\vec q)=G_\pi(-\vec q)$ in isotropic nuclear matter. From now on, we omit the momentum dependence of the form factors for simplicity. It is a straightforward task to arrive at a similar expression for the neutral pion, i.e., $a=b=0$,
\begin{align}
    \Pi^{00} (q) &= \im \sum_{n^0}\bigg[ \frac{G^{*}_{\pi^0_n} \bar G^*_{\pi^0_n}}{2\omega_{n^0}(\vec q)(q^0 -\omega_{n^0}(\vec q) + \im \epsilon)} -  \frac{\bar G^{*}_{\pi^0_n} G^*_{\pi^0_n}}{2\omega_{n^0}(\vec q)(q^0 + \omega_{n^0}(\vec q) - \im \epsilon)}\bigg].
\end{align}

In similar ways, we can obtain the expressions for other correlation functions: 
\begin{align}
    \im q^\mu\Pi_{5\mu}^{\pm\mp} (q) &= -\im q^\mu \left[ \sum_{n^\pm} \frac{(n_\mu N_{\pi_n^\pm}^*+p_\mu^{n^\pm}(\vec q) F^*_{\pi_n^\pm}) \bar G^*_{\pi^\pm_n}}{2\omega_{n^\pm}(\vec q)\big(q^0 -\omega_{n^\pm}(\vec q) + \im \epsilon\big)} + \sum_{n^\mp} \frac{(n_\mu \bar N_{\pi_n^\mp}^*+p_\mu^{n^\mp}(-\vec q) \bar F^*_{\pi_n^\mp}) G^*_{\pi^\mp_n}}{2\omega_{n^\mp}(\vec q)\big(q^0 + \omega_{n^\mp}(\vec q) - \im \epsilon\big)} \right]\\
    \im q^\mu\Pi_{5\mu}^{00} (q) &= -\im q^\mu \sum_{n^0}\left[ \frac{(n_\mu N_{\pi_n^0}^*+p_\mu^{n^0}(\vec q) F^*_{\pi_n^0}) \bar G^*_{\pi^0_n}}{2\omega_{n^0}(\vec q)\big(q^0 -\omega_{n^0}(\vec q) + \im \epsilon\big)} + \frac{(n_\mu \bar N_{\pi_n^0}^*+p_\mu^{n^0}(-\vec q) \bar F^*_{\pi_n^0}) G^*_{\pi^0_n}}{2\omega_{n^0}(\vec q)\big(q^0 + \omega_{n^0}(\vec q) - \im \epsilon\big)} \right]
\end{align}
\begin{align}
    \Pi_{\mu\nu}^{\pm\mp}(q)&= \im \bigg[ \sum_{n^\pm} \frac{(n_\mu N_{\pi_n^\pm}^*+p_\mu^{n^\pm}(\vec q) F^*_{\pi_n^\pm})(n_\nu \bar N_{\pi_n^\pm}^*+p_\nu^{n^\pm}(\vec q) \bar F^*_{\pi_n^\pm})}{2\omega_{n^\pm}(\vec q)\big(q^0 -\omega_{n^\pm}(\vec q) + \im \epsilon\big)}- \sum_{n^\mp} \frac{(n_\mu \bar N_{\pi_n^\mp}^*+p_\mu^{n^\mp}(-\vec q) \bar F^*_{\pi_n^\mp})(n_\nu N_{\pi_n^\mp}^*+p_\nu^{n^\mp}(-\vec q)  F^*_{\pi_n^\mp})}{2\omega_{n^\mp}(\vec q)\big(q^0 + \omega_{n^\mp}(\vec q) - \im \epsilon\big)} \bigg]\\
    \Pi_{\mu\nu}^{00}(q)&= \im \sum_{n^0}\bigg[ \frac{(n_\mu N_{\pi_n^0}^*+p_\mu^{n^0}(\vec q) F^*_{\pi_n^0})(n_\nu \bar N_{\pi_n^0}^*+p_\nu^{n^0}(\vec q) \bar F^*_{\pi_n^0})}{2\omega_{n^0}(\vec q)\big(q^0 -\omega_{n^0}(\vec q) + \im \epsilon\big)}- \frac{(n_\mu \bar N_{\pi_n^0}^*+p_\mu^{n^0}(-\vec q) \bar F^*_{\pi_n^0})(n_\nu N_{\pi_n^0}^*+p_\nu^{n^0}(-\vec q)  F^*_{\pi_n^0})}{2\omega_{n^0}(\vec q)\big(q^0 + \omega_{n^0}(\vec q) - \im \epsilon\big)} \bigg],
\end{align}
where $p^\mu_n(\vec q)=(\omega_n(\vec q),\vec q)$ and $F_{\pi^n}(\vec q)\equiv F_{\pi^n}(\omega_n(\vec q),\vec q)$ for the form factors which have symmetry under $\vec q\leftrightarrow-\vec q$ in isotropic matter. In this paper, we mainly focus on the lowest pion states for the hadronic states $n^{\pm,0}$.

\end{widetext}
\section{Derivation of the in-medium Gell-Mann-Oakes-Renner relation}\label{app:gor}
The Gell-Mann-Oakes-Renner relation in nuclear matter, which relates the pion properties to the quark condensate, is considered with the nonzero quark mass. To derive this relation, we start with the following correlation function,
\begin{widetext}
\begin{align}\label{eq:corr-function}
	& \int\dd[4]{x} \eu^{\im q\cdot x} \partial^\mu \langle\Omega|\mathsf T A_\mu^{\pm,0} (x) P^{\mp,0}(0)|\Omega\rangle=-\im q^\mu\int\dd[4]{x} \eu^{\im q\cdot x} \langle\Omega|\mathsf T A_\mu^{\pm,0} (x) P^{\mp,0}(0)|\Omega\rangle\nonumber \\
	&= \int\dd[4]{x} \eu^{\im q\cdot x} \langle\Omega|\mathsf T \partial^\mu A_\mu^{\pm,0} (x) P^{\mp,0}(0)|\Omega\rangle+\int\dd[4]{x} \eu^{\im q\cdot x} \delta(x_0)\langle\Omega|[ A_0^{\pm,0} (x), P^{\mp,0}(0)]|\Omega\rangle\nonumber \\
    &= m_q\int\dd[4]{x} \eu^{\im q\cdot x} \langle\Omega|\mathsf T P^{\pm,0} (x) P^{\mp,0}(0)|\Omega\rangle+\int\dd[4]{x} \eu^{\im q\cdot x} \delta(x_0)\langle\Omega|[ A_0^{\pm,0} (x), P^{\mp,0}(0)]|\Omega\rangle ,
\end{align}
where we have used the integration by parts in the first line and the partially conserved axial current (PCAC) relation $\partial^\mu A_\mu^{\pm,0}(x)=m_qP^{\pm,0}(x)$ in the third line. 
With the abbreviated expressions, we arrive at
\begin{align}\label{eq:gor1}
    \im\lim_{\vec q \to 0}q^\mu\Pi^{\pm\mp,00}_{5\mu}(q)+m_q\lim_{\vec q\rightarrow0}\Pi^{\pm\mp,00}(q)&=-\langle\Omega|[Q_5^{\pm,0},P^{\mp,0}(0)]|\Omega\rangle\nonumber\\
    &=\im\langle\bar qq\rangle^*,
\end{align}
where we used
\begin{align}
    Q_5^{\pm,0}=\int \dd[3]x A_0^{\pm,0}(x),\qquad [Q_5^{\pm,0},P^{\mp,0}]=-\im\bar qq.
\end{align}

We consider the correlation functions with $\mu=0$. Then, the correlation functions using the hadronic quantities can be written as in the limit of $\vec q\rightarrow0$:
\begin{subequations}\label{eq:in-medium-GWNF}
\begin{align}
    m_q\lim_{\vec q \to 0}\Pi^{00} (q) &= \im m_q \sum_{n^0} \left[ \frac{G^{*}_{\pi^0_n} \bar G^*_{\pi^0_n}}{2m_{\pi_n^0}^*\left(q^0 -m_{\pi_n^0}^* + \im \epsilon\right)} - \frac{\bar G^{*}_{\pi^0_n} G^*_{\pi^0_n}}{2m_{\pi_n^0}^*\left(q^0 + m_{\pi_n^0}^* - \im \epsilon\right)} \right],\\
    m_q\lim_{\vec q \to 0}\Pi^{\pm\mp} (q) &= \im m_q \left[ \sum_{n^\pm} \frac{G^{*}_{\pi^\pm_n} \bar G^*_{\pi^\pm_n}}{2m_{\pi_n^\pm}^*\left(q^0 -m_{\pi_n^\pm}^* + \im \epsilon\right)} - \sum_{n^\mp}\frac{\bar G^{*}_{\pi^\mp_n} G^*_{\pi^\mp_n}}{2m_{\pi_n^\mp}^*\left(q^0 + m_{\pi_n^\mp}^* - \im \epsilon\right)} \right],\\
    \im q^0\lim_{\vec q \to 0}\Pi_{50}^{00} (q) &= -\im q^0 \sum_{n^0} \left[ \frac{\left(m^*_{\pi_n^0}F^*_{\pi_n^0}+N_{\pi_n^0}^*\right) \bar G^*_{\pi^0_n}}{2m^*_{\pi_n^0}\left(q^0 -m_{\pi_n^0}^* + \im \epsilon\right)} + \frac{\left(m^*_{\pi_n^0}\bar{F}^*_{\pi_n^0}+\bar{N}_{\pi_n^0}^*\right) G^*_{\pi^0_n}}{2m_{\pi_n^0}^*\left(q^0 + m_{\pi_n^0}^* - \im \epsilon\right)} \right],\\
    \im q^0\lim_{\vec q \to 0}\Pi_{50}^{\pm\mp} (q) &= -\im q^0\left[\sum_{n^\pm} \frac{\left(m^*_{\pi_n^\pm}F^*_{\pi_n^\pm}+N_{\pi_n^\pm}^*\right) \bar G^*_{\pi^\pm_n}}{2m_{\pi_n^\pm}^*\left(q^0 -m_{\pi_n^\pm}^* + \im \epsilon\right)} + \sum_{n^\mp} \frac{\left(m^*_{\pi_n^\mp}\bar{F}^*_{\pi_n^\mp}+\bar{N}_{\pi_n^\mp}^*\right) G^*_{\pi^\mp_n}}{2m_{\pi_n^\mp}^*\left(q^0 + m_{\pi_n^\mp}^* - \im \epsilon\right)} \right],
\end{align}
\end{subequations}
where $n^{\pm,0}$ label $\pi^{\pm,0}$ excited states.
We now use the PCAC relation for the pion states \cite{Jido2008}
\begin{align}    
    (\omega_{n^{\pm,0}}(\vec p)^2- \vec p \cdot \vec p )F_{\pi_n^{\pm,0}}^{*}+\omega_{n^{\pm,0}}(\vec p)N_{\pi_n^{\pm,0}}^{*}  = m_q G_{\pi_n^{\pm,0}}^*.
\end{align}
In the rest frame, then, we arrive at
\begin{subequations}\label{eq:gorfn}
\begin{align}
    \frac{1}{2} \sum_{n^+}  \left[\left|m^*_{\pi_n^+}F^*_{\pi_n^+}+N_{\pi_n^+}^*\right|^2\right] 
    + \frac{1}{2} \sum_{n^-}  \left[\left|m^*_{\pi_n^-}F^*_{\pi_n^-}+N_{\pi_n^-}^*\right|^2 \right]&=-m_q\langle \bar qq\rangle^*,\label{eq:apqqch}\\
    \sum_{n^0}  \left[\left|m^*_{\pi_n^0}F^*_{\pi_n^0}+N_{\pi_n^0}^*\right|^2 \right]&=-m_q\langle \bar qq\rangle^*.\label{eq:apqqn}
\end{align}
\end{subequations}
Equation (\ref{eq:apqqch}) has been obtained from the correlation functions for the charged particles, which provide the equivalent sum rule, while the correlation functions for the neutral pion give \cref{eq:apqqn}. Both sum rules share the same right-hand side. These are the generalized GOR relations in isospin asymmetric nuclear matter. It can be considered in terms of the decay constants.
By using the definition of the temporal decay constant given as \cref{eq:chfn}, we can write \cref{eq:gorfn} to
\begin{subequations}\label{eq:gor3}
\begin{align}
    \frac{1}{2} \sum_{n^+}  \left[ m_{\pi_n^+}^{*2}\left|f^{*(t)}_{\pi^+_n}\right|^2 \right] 
    + \frac{1}{2} \sum_{n^-}  \left[ m_{\pi_n^-}^{*2}\left|f^{*(t)}_{\pi^-_n}\right|^2 \right]&=-m_q\langle \bar qq\rangle^*,\label{eq:qqcharged}\\
    \sum_{n^0}  \left[ m_{\pi_n^0}^{*2}\left|f^{*(t)}_{\pi^0_n}\right|^2 \right]&=-m_q\langle \bar qq\rangle^*.
\end{align}
\end{subequations}
\end{widetext}
Note that the final results are independent of arbitrary $q^0$, and all pion excited states are related to the quark condensate. Although one considers the $+-$ and $-+$ components of the correlation functions separately, \cref{eq:qqcharged} can be obtained.

\section{Interaction terms}\label{app:interactions}
In this appendix, we list all the interactions in the Lagrangian that we use in this work. 
The particles that interact in the corresponding vertex are indicated via subscripts. 
First, these are the interactions coming from the chiral Lagrangian in \cref{eq:pionLag}, involving only the pion field and external sources:
\begin{subequations}
\begin{align}
    \mathcal L_{\pi^4}^{(2)} &= \frac{1}{10f^2}\partial_\mu \pi^i \partial^\mu \pi^j \pi^k\pi^l (3\delta^{ik}\delta^{jl}-\delta^{ij}\delta^{kl}) \nonumber\\
    &\qquad -\frac{m B_0}{20f^2}\pi^i\pi^i\pi^j\pi^j, \\
    \mathcal L_{\pi a}^{(2)} &= -f \partial^\mu \pi^i a^j_\mu \delta^{ij}, \\ 
    \mathcal L_{\pi^3 a}^{(2)} &= \frac{1}{5f} a^i_\mu \partial^\mu \pi^j \pi^k\pi^l (3\delta^{ij}\delta^{kl}-4\delta^{ik}\delta^{jl}).
\end{align}
\end{subequations}
Using the first term in the expansion of the pion-nucleon interaction operator $A$ in \cref{eq:A1}, we find the following interactions relevant for this work (note that the only parameter is $g_A$, the rest is fixed by chiral symmetry; also additionally to the fields indicated in the subscript, a nucleon enters and exits the vertex):
\begin{subequations}
\begin{align}
	A_{\pi}^{(1)} &= \frac{g_A}{2f}\gamma^\mu\gamma^5 \partial_\mu \pi^i \tau^i, \\
	A_{\pi\pi}^{(1)} &= \frac{1}{4f^2} \gamma^\mu \pi^i \partial_\mu \pi^j \epsilon^{ijk} \tau^k, \\
	A_{\pi^3}^{(1)} &= \frac{g_A}{20f^3} \gamma^\mu\gamma^5 \left[ 3\pi^i \pi^j \partial_\mu\pi^j - \pi^j\pi^j \partial_\mu\pi^i \right]\tau^i, \\
	A_{a}^{(1)} &= -g_A \gamma^\mu\gamma^5 a_\mu^i \frac{\tau^i}{2}, \\
	A_{\pi a}^{(1)} &= -\frac{1}{2f}\gamma^\mu \pi^ia_\mu^j \epsilon^{ijk}\tau^k, \\
	A_{\pi^2 a}^{(1)} &= \frac{g_A}{4f^2} \gamma^\mu \gamma^5 \pi^a \pi^b a_\mu^c \tau^d (\delta^{ab}_{cd} - \delta^{ac}_{bd}). 
\end{align}
\end{subequations}
From the second term of the pion-nucleon interaction operator, $A^{(2)}$ in \cref{eq:A2}, we get these interactions:
\begin{subequations}
\begin{align}
    A_{\pi\pi}^{(2)} &= \frac{4B_0 c_1 m}{f^2} \pi^i\pi^i + \frac{c_2}{f^2m_N^2}\partial_\mu \pi^i \partial_\nu \pi^i \partial^\mu\partial^\nu \nonumber\\
    &\qquad - \frac{c_3}{f^2} \partial_\mu \pi^i \partial^\mu \pi^i + \frac{\im c_4}{f^2}\epsilon^{ijk}\tau^k \gamma^\mu \gamma^\nu \partial_\mu \pi^i \partial_\nu \pi^j, \\
    A_{\pi a}^{(2)} &= -\frac{2c_2}{fm_N^2}\partial_\mu \pi^i a_\nu ^i \partial^\mu \partial^\nu + \frac{2c_3}{f}\partial^\mu \pi^i a^i_\mu \nonumber\\
    &\qquad - \frac{\im c_4}{f} \epsilon^{ijk}\tau^k \partial_\mu \pi^i a_\nu^j [\gamma^\mu,\gamma^\nu], 
\end{align}
\end{subequations}
and we also used the fact that the following vertices do not exist: $A_{\pi}^{(2)} = 0$, $A_{\pi^3}^{(2)} = 0$. 

\section{Interaction terms in the physical basis}\label{app:interactions_physical}
In our calculations of the self-energy and decay constant, we use the interaction terms represented by the physical basis, $\{ \pi^+,\pi^-,\pi^0 \}$. These are obtained by applying the transformation rules, 
\begin{align}
\pi^3 &= \pi^0, \\ 
\pi^1 &= \frac{1}{\sqrt{2}}(\pi^+ + \pi^-),\ \pi^2 = \frac{\im}{\sqrt{2}} (\pi^+ - \pi^-),
\end{align}
and 
\begin{align}
a_\mu^3 &= a_\mu^0,\\
a_\mu^1 &= \frac{1}{\sqrt{2}}(a_\mu^+ + a_\mu^-),\ a_\mu^2 = \frac{\im}{\sqrt{2}} (a_\mu^+ - a_\mu^-),
\end{align}
to the expressions given in the previous section. We present their explicit forms in the following:
\begin{widetext}
\begin{subequations}
\begin{align}
    \mathcal L_{\pi^4}^{(2)} 
    &= \frac{1}{10f^2} \left( 
    2 \partial_\mu \pi^+ \partial^\mu \pi^- \pi^+ \pi^-
    - 2 \partial_\mu \pi^+ \partial^\mu \pi^- \pi^0 \pi^0
    - 2 \partial_\mu \pi^0 \partial^\mu \pi^0 \pi^+ \pi^-
    + 2 \partial_\mu \pi^0 \partial^\mu \pi^0 \pi^0 \pi^0 
    \right. \nonumber \\ 
    &\qquad \left. 
    + 3 \partial_\mu \pi^+ \partial^\mu \pi^+ \pi^- \pi^-
    + 3 \partial_\mu \pi^- \partial^\mu \pi^- \pi^+ \pi^+
    + 6 \partial_\mu \pi^+ \partial^\mu \pi^0 \pi^- \pi^0
    + 6 \partial_\mu \pi^-    \partial^\mu \pi^0 \pi^+ \pi^0
    \right) \nonumber \\ 
    &\qquad
    - \frac{B_0 m_q}{20f^2} 
    (
    4 \pi^+ \pi^- \pi^+ \pi^-
    + 4 \pi^+ \pi^- \pi^0 \pi^0
    + \pi^0 \pi^0 \pi^0 \pi^0
    ), \\
    \mathcal L_{\pi a}^{(2)} &= -f (\partial^\mu \pi^+ a_\mu^- + \partial^\mu \pi^- a^+_\mu + \partial \pi^0 a^0_\mu), \\ 
    \mathcal L_{\pi^3 a}^{(2)} 
    &= \frac{1}{5f} \left(
    2 a_\mu^+ \partial^\mu \pi^- \pi^+ \pi^-
    + 3 a_\mu^+ \partial^\mu \pi^- \pi^0 \pi^0
    - 4 a_\mu^+ \partial^\mu \pi^+ \pi^- \pi^-
    - 4 a_\mu^+ \partial^\mu \pi^0 \pi^0 \pi^-
    \right. \nonumber \\ 
    &\qquad \left. 
    + 2 a_\mu^- \partial^\mu \pi^+ \pi^- \pi^+
    + 3 a_\mu^- \partial^\mu \pi^+ \pi^0 \pi^0
    - 4 a_\mu^- \partial^\mu \pi^- \pi^+ \pi^+
    - 4 a_\mu^- \partial^\mu \pi^0 \pi^0 \pi^+
    \right. \nonumber \\ 
    &\qquad \left. 
    - a_\mu^0 \partial^\mu \pi^0 \pi^0 \pi^0
        + 6 a_\mu^0 \partial^\mu \pi^0 \pi^+ \pi^-
        - 4 a_\mu^0 \partial^\mu \pi^- \pi^+ \pi^0
        - 4 a_\mu^0 \partial^\mu \pi^+ \pi^- \pi^0
    \right).
\end{align}
\end{subequations}
\begin{subequations}
\begin{align}
    A_{\pi}^{(1)} &= 
    \frac{g_A}{2f} \gamma^\mu \gamma^5 
    \left(
    \partial_\mu \pi^0 \tau^3
    + 
    \sqrt{2} \partial_\mu \pi^+ \tau^+
    +
    \sqrt{2} \partial_\mu \pi^- \tau^-
    \right), \\
    A_{\pi\pi}^{(1)} &= 
    \frac{\gamma^\mu}{4 \im f^2} \left[
    (\pi^+ \partial_\mu \pi^- - \pi^- \partial_\mu \pi^+) \tau^3
    +
    (\sqrt{2}\pi^0 \partial_\mu \pi^+ - \sqrt{2}\pi^+ \partial_\mu \pi^0) \tau^+
    +
    (\sqrt{2}\pi^- \partial_\mu \pi^0 - \sqrt{2}\pi^0 \partial_\mu \pi^-) \tau^-
    \right] 
    , \\
    A_{\pi^3}^{(1)} &= 
    \frac{g_A}{20f^3} \gamma^\mu\gamma^5 
    \left( 
    3 \pi^0 \pi^- \partial_\mu \pi^+ 
    + 3 \pi^0 \pi^+ \partial_\mu \pi^-
    + 2 \pi^0 \pi^0 \partial_\mu \pi^0
    - 2 \pi^- \pi^+ \partial_\mu \pi^0
    \right)\tau^3 \nonumber \\
    &\qquad + \frac{\sqrt{2}g_A}{20f^3} \gamma^\mu\gamma^5
    \left( 
    \pi^+ \pi^- \partial_\mu \pi^+
    + 3 \pi^+ \pi^+ \partial_\mu \pi^-
    + 3 \pi^+ \pi^0 \partial_\mu \pi^0
    - \pi^0 \pi^0 \partial_\mu \pi^+
    \right)\tau^+ \nonumber \\
    &\qquad + \frac{\sqrt{2}g_A}{20f^3} \gamma^\mu\gamma^5
    \left( 
    \pi^+ \pi^- \partial_\mu \pi^-
    + 3 \pi^- \pi^- \partial_\mu \pi^+
    + 3 \pi^- \pi^0 \partial_\mu \pi^0
    - \pi^0 \pi^0 \partial_\mu \pi^-
    \right)\tau^-, \\ 
    A_{a}^{(1)} &= 
    - \frac{g_A}{2} \gamma^\mu\gamma^5 
    \left( 
    a_\mu^0 \tau^3 + \sqrt{2} a_\mu^+ \tau^+ + \sqrt{2} a_\mu^- \tau^-
    \right), \\
    A_{\pi a}^{(1)} &= 
    -\frac{\im}{2f} \gamma^\mu 
    \left[
    (a_\mu^+ \pi^- - a_\mu^- \pi^+) \tau^3 
    + \sqrt{2} (a_\mu^0 \pi^+ - a_\mu^+ \pi^0) \tau^+ 
    + \sqrt{2} (a_\mu^- \pi^0 - a_\mu^0 \pi^-) \tau^-
    \right], \\
    A_{\pi^2 a}^{(1)} &= 
    \frac{g_A}{4f^2} \gamma^\mu \gamma^5 
    \left[ 
    (2 a_\mu^0 \pi^+ \pi^- - a_\mu^+ \pi^- \pi^0 - a_\mu^- \pi^+ \pi^0) \tau^3 \right. \nonumber \\ 
    &\qquad \left. 
    + \sqrt{2} 
    (a_\mu^+ \pi^+ \pi^- - a_\mu^- \pi^+ \pi^+ - a_\mu^0 \pi^0 \pi^+ + a_\mu^+ \pi^0 \pi^0) \tau^+ 
    \right. \nonumber \\ 
    &\qquad \left. 
    + \sqrt{2} 
    (-a_\mu^+ \pi^- \pi^- + a_\mu^- \pi^+ \pi^- - a_\mu^0 \pi^0 \pi^- + a_\mu^- \pi^0 \pi^0) \tau^- 
    \right]. 
\end{align}
\end{subequations}
where we define $\tau^\pm = (\tau^1 \pm \im \tau^2)/2$ such that $[\tau^+, \tau^-]=\tau^3$. From the second term of the pion-nucleon interaction operator, $A^{(2)}$, we get these interactions:
\begin{subequations}
\begin{align}
    A_{\pi\pi}^{(2)} &= 
    \frac{4B_0 c_1 m}{f^2} (2 \pi^+ \pi^- + \pi^0 \pi^0) 
    + \frac{c_2}{f^2m_N^2} (
    \partial_\mu \pi^+ \partial_\nu \pi^-
    + \partial_\mu \pi^- \partial_\nu \pi^+
    + \partial_\mu \pi^0 \partial_\nu \pi^0
    ) \partial^\mu\partial^\nu \nonumber\\
    &\qquad - \frac{c_3}{f^2} 
    (2 \partial_\mu \pi^+ \partial^\mu \pi^- 
    + \partial_\mu \pi^0 \partial^\mu \pi^0)
    + \frac{c_4}{f^2} \gamma^\mu \gamma^\nu 
    (\partial_\mu \pi^+ \partial_\nu \pi^- - \partial_\mu \pi^- \partial_\nu \pi^+) \tau^3 \nonumber \\ 
    &\qquad 
    + \frac{c_4}{f^2} \gamma^\mu \gamma^\nu \left[ 
    (\sqrt{2}\partial_\mu \pi^0 \partial_\nu \pi^+ - \sqrt{2}\partial_\mu \pi^+ \partial_\nu \pi^0) \tau^+ 
    + (\sqrt{2}\partial_\mu \pi^- \partial_\nu \pi^0 - \sqrt{2}\partial_\mu \pi^0 \partial_\nu \pi^-) \tau^- 
    \right], \\
    A_{\pi a}^{(2)} &= 
    \partial_\mu \pi^+ a_\nu^- \left( - \frac{2c_2}{fm_N^2} \partial^\mu \partial^\nu - \frac{c_4}{f} \tau^3 [\gamma^\mu, \gamma^\nu] \right) + \frac{2c_3}{f} \partial^\mu \pi^+ a_\mu^- \nonumber \\ 
    &\qquad + \partial_\mu \pi^- a_\nu^+ \left( - \frac{2c_2}{fm_N^2} \partial^\mu \partial^\nu + \frac{c_4}{f} \tau^3 [\gamma^\mu, \gamma^\nu] \right) + \frac{2c_3}{f} \partial^\mu \pi^- a_\mu^+ 
    - \frac{2c_2}{fm_N^2} \partial_\mu \pi^0 a_\nu^0 \partial^\mu \partial^\nu + \frac{2c_3}{f} \partial^\mu \pi^0 a_\mu^0 \nonumber \\ 
    &\qquad 
    + \frac{c_4}{f} \left( \sqrt{2} \partial_\mu \pi^+ a_\nu^0 - \sqrt{2} \partial_\mu \pi^0 a_\nu^+ \right) \tau^+ [\gamma^\mu, \gamma^\nu]
    - \frac{c_4}{f} \left( \sqrt{2} \partial_\mu \pi^- a_\nu^0 - \sqrt{2} \partial_\mu \pi^0 a_\nu^- \right) \tau^- [\gamma^\mu, \gamma^\nu].
\end{align}
\end{subequations}
\end{widetext}
\section{Explicit forms of the self-energies and the in-medium vertex corrections}\label{app:Selfenergy_Decayconst}
We display explicit forms of the Feynman diagrams \cref{self-energy diagrams,axial 1PI diagrams} in this appendix.
First, the expressions of the one-loop contributions of the self-energies of the in-medium pions are given by
\begin{widetext}
\begin{subequations}\label{eq:result-self-energy}\allowdisplaybreaks
\begin{align}
    \Sigma_{\mathrm{1\text{-}loop}}^{+}(q_0)
    &= \int \dd |\vec p| \frac{\vec p^2}{4\pi^2p_0f^2} \left[ \frac{2q_0g_A^2(q_0p_0+2\vec p^2)}{q_0+2p_0}\Theta^n_{\boldsymbol{p}}-\frac{2q_0g_A^2(q_0p_0-2\vec p^2)}{q_0-2p_0}\Theta^p_{\boldsymbol{p}}\right. \nonumber\\
    &\qquad \left. +4m_N\left(4c_1m_\pi^2-\frac{2c_2}{m_N^2}q_0^2p_0^2-2c_3q_0^2\right)(\Theta^p_{\boldsymbol{p}}+\Theta^n_{\boldsymbol{p}})+2q_0p_0(\Theta^p_{\boldsymbol{p}}-\Theta^n_{\boldsymbol{p}})\right] \nonumber\\
    &\approx\frac{q_0^2g_A^2m_N\rho}{f^2(4m_N^2-q_0^2)}+\frac{q_0^3g_A^2(1-r)\rho}{2f^2(1+r)(4m_N^2-q_0^2)}+\frac{[4c_1m_\pi^2-2q_0^2(c_2+c_3)]\rho}{f^2}+\frac{q_0(1-r)\rho}{2f^2(1+r)},\\ \nonumber \\ 
    \Sigma_{\mathrm{1\text{-}loop}}^{-}(q_0)
    &=\int \dd |\vec p|\frac{\vec p^2}{4\pi^2p_0f^2}\left[\frac{2q_0g_A^2(q_0p_0+2\vec p^2)}{q_0+2p_0}\Theta^p_{\boldsymbol{p}}-\frac{2q_0g_A^2(q_0p_0-2\vec p^2)}{q_0-2p_0}\Theta^n_{\boldsymbol{p}} \right. \nonumber \\
    &\qquad \left. +4m_N\left(4c_1m_\pi^2-\frac{2c_2}{m_N^2}q_0^2p_0^2-2c_3q_0^2\right)(\Theta^p_{\boldsymbol{p}}+\Theta^n_{\boldsymbol{p}})-2q_0p_0(\Theta^p_{\boldsymbol{p}}-\Theta^n_{\boldsymbol{p}})\right]\nonumber\\
    &\approx\frac{q_0^2g_A^2m_N\rho}{f^2(4m_N^2-q_0^2)}-\frac{q_0^3g_A^2(1-r)\rho}{2f^2(1+r)(4m_N^2-q_0^2)}+\frac{[4c_1m_\pi^2-2q_0^2(c_2+c_3)]\rho}{f^2}-\frac{q_0(1-r)\rho}{2f^2(1+r)},\\ \nonumber \\ 
    \Sigma_{\mathrm{1\text{-}loop}}^{0}(q_0)
    &=\int \dd |\vec p|\frac{\vec p^2}{8\pi^2p_0f^2}\left[\frac{2q_0g_A^2(q_0p_0+2\vec p^2)}{q_0+2p_0}(\Theta^p_{\boldsymbol{p}}+\Theta^n_{\boldsymbol{p}})-\frac{2q_0g_A^2(q_0p_0-2\vec p^2)}{q_0-2p_0}(\Theta^n_{\boldsymbol{p}}+\Theta^p_{\boldsymbol{p}})\right. \nonumber \\
    &\qquad \left. +8m_N\left(4c_1m_\pi^2-\frac{2c_2}{m_N^2}q_0^2p_0^2-2c_3q_0^2\right)(\Theta^p_{\boldsymbol{p}}+\Theta^n_{\boldsymbol{p}})\right]\nonumber\\
    &\approx \frac{q_0^2g_A^2m_N\rho}{f^2(4m_N^2-q_0^2)}+\frac{[4c_1m_\pi^2-2q_0^2(c_2+c_3)]\rho}{f^2}.
\end{align}
where we abbreviate the neutron-to-proton ratio $r\equiv \rho_n/\rho_p$ and the total density as $\rho=\rho_n+\rho_p$.
We represent linear contributions of the nuclear density for the one-loop contributions of each pion, which are shown on the right-hand side of $\approx$. The two-loop contributions are given by
\begin{align}
    \Sigma_{\mathrm{2\text{-}loop}}^{+}(q_0)
    &=\frac{m_N^2g_A^2}{10(2\pi f)^4}\int \dd |\vec p|\,\dd |\vec k|\,\dd \cos \theta\frac{\vec p^2\vec k^2}{p_0k_0}\frac{p_0k_0-|\vec p||\vec k|\cos\theta-m_N^2}{(2m_N^2-m_\pi^2-2p_0k_0+2|\vec p||\vec k|\cos\theta)^2}\nonumber\\
    &\times[(20q_0p_0-20q_0k_0-4p_0k_0+4|\vec p||\vec k|\cos\theta+4m_N^2+2q_0^2-4m_\pi^2)\Theta_{\boldsymbol{k}}^p\Theta_{\boldsymbol{p}}^n\nonumber\\
    &+(-20q_0p_0+20q_0k_0-4p_0k_0+4|\vec p||\vec k|\cos\theta+4m_N^2+2q_0^2-4m_\pi^2)\Theta_{\boldsymbol{k}}^n\Theta_{\boldsymbol{p}}^p\nonumber\\
    &+(4p_0k_0-4|\vec p||\vec k|\cos\theta-4m_N^2-2q_0^2-m_\pi^2)(\Theta_{\boldsymbol{k}}^p\Theta_{\boldsymbol{p}}^p+\Theta_{\boldsymbol{k}}^n\Theta_{\boldsymbol{p}}^n)]+\frac{m_N^2g_A^2}{10(2\pi f)^4}\int \dd |\vec p|\,\dd |\vec k|\,\dd \cos \theta\frac{\vec p^2\vec k^2}{p_0k_0}\nonumber\\
    &\times\frac{1}{2m_N^2-m_\pi^2-2p_0k_0+2|\vec p||\vec k|\cos\theta}[(5q_0k_0-5q_0p_0+2m_N^2-2p_0k_0+2|\vec p||\vec k|\cos\theta)\Theta^p_{\boldsymbol{p}}\Theta_{\boldsymbol{k}}^n\nonumber\\
    &+(-5q_0k_0+5q_0p_0+2m_N^2-2p_0k_0+2|\vec p||\vec k|\cos\theta)\Theta^p_{\boldsymbol{k}}\Theta_{\boldsymbol{p}}^n+(2p_0k_0-2|\vec p||\vec k|\cos\theta-2m_N^2)(\Theta^p_{\boldsymbol{p}}\Theta^p_{\boldsymbol{k}}+\Theta^n_{\boldsymbol{p}}\Theta^n_{\boldsymbol{k}})]\nonumber\\
    &-\frac{q_0^2}{4(2\pi)^4f^4}\int \dd |\vec p|\,\dd |\vec k|\,\dd \cos \theta\frac{\vec p^2\vec k^2}{p_0k_0}\frac{p_0k_0+|\vec p||\vec k|\cos\theta+m_N^2}{(q+p-k)^2-m_\pi^2}(\Theta^p_{\boldsymbol{p}}\Theta^p_{\boldsymbol{k}}+\Theta^n_{\boldsymbol{k}}\Theta^n_{\boldsymbol{p}}+2\Theta^n_{\boldsymbol{p}}\Theta^p_{\boldsymbol{k}})\nonumber\\
    &-\frac{1}{(2\pi)^4}\int \dd |\vec p|\,\dd |\vec k|\,\dd \cos \theta\frac{\vec p^2\vec k^2}{p_0k_0}\frac{C_L^sC_R^s(p_0k_0-|\vec p||\vec k|\cos\theta+m_N^2)}{(q+p-k)^2-m_\pi^2}(\Theta^p_{\boldsymbol{p}}\Theta^p_{\boldsymbol{k}}+\Theta^n_{\boldsymbol{k}}\Theta^n_{\boldsymbol{p}})\nonumber\\
    &-\frac{m_Nq_0}{2(2\pi)^4f^2}\int \dd |\vec p|\,\dd |\vec k|\,\dd \cos \theta\frac{\vec p^2\vec k^2}{p_0k_0}\frac{(p_0+k_0)}{(q+p-k)^2-m_\pi^2}(\Theta^p_{\boldsymbol{p}}\Theta^p_{\boldsymbol{k}}-\Theta^n_{\boldsymbol{k}}\Theta^n_{\boldsymbol{p}})(C_L^s+C_R^s), 
\end{align}
\begin{align}
    \Sigma_{\mathrm{2\text{-}loop}}^{-}(q_0)
    &=\frac{m_N^2g_A^2}{10(2\pi f)^4}\int \dd |\vec p|\,\dd |\vec k|\,\dd \cos \theta\frac{\vec p^2\vec k^2}{p_0k_0}\frac{p_0k_0-|\vec p||\vec k|\cos\theta-m_N^2}{(2m_N^2-m_\pi^2-2p_0k_0+2|\vec p||\vec k|\cos\theta)^2}\nonumber\\
    &\times[(-20q_0p_0+20q_0k_0-4p_0k_0+4|\vec p||\vec k|\cos\theta+4m_N^2+2q_0^2-4m_\pi^2)\Theta_{\boldsymbol{k}}^p\Theta_{\boldsymbol{p}}^n\nonumber\\
    &+(20q_0p_0-20q_0k_0-4p_0k_0+4|\vec p||\vec k|\cos\theta+4m_N^2+2q_0^2-4m_\pi^2)\Theta_{\boldsymbol{k}}^n\Theta_{\boldsymbol{p}}^p\nonumber\\
    &+(4p_0k_0-4|\vec p||\vec k|\cos\theta-4m_N^2-2q_0^2-m_\pi^2)(\Theta_{\boldsymbol{k}}^p\Theta_{\boldsymbol{p}}^p+\Theta_{\boldsymbol{k}}^n\Theta_{\boldsymbol{p}}^n)]+\frac{m_N^2g_A^2}{10(2\pi f)^4}\int \dd |\vec p|\,\dd |\vec k|\,\dd \cos \theta\frac{\vec p^2\vec k^2}{p_0k_0}\nonumber\\
    &\times\frac{1}{2m_N^2-m_\pi^2-2p_0k_0+2|\vec p||\vec k|\cos\theta}[(-5q_0k_0+5q_0p_0+2m_N^2-2p_0k_0+2|\vec p||\vec k|\cos\theta)\Theta^p_{\boldsymbol{p}}\Theta_{\boldsymbol{k}}^n\nonumber\\
    &+(5q_0k_0-5q_0p_0+2m_N^2-2p_0k_0+2|\vec p||\vec k|\cos\theta)\Theta^p_{\boldsymbol{k}}\Theta_{\boldsymbol{p}}^n+(2p_0k_0-2|\vec p||\vec k|\cos\theta-2m_N^2)(\Theta^p_{\boldsymbol{p}}\Theta^p_{\boldsymbol{k}}+\Theta^n_{\boldsymbol{p}}\Theta^n_{\boldsymbol{k}})]\nonumber\\
    &-\frac{q_0^2}{4(2\pi)^4f^4}\int \dd |\vec p|\,\dd |\vec k|\,\dd \cos \theta\frac{\vec p^2\vec k^2}{p_0k_0}\frac{p_0k_0+|\vec p||\vec k|\cos\theta+m_N^2}{(q+p-k)^2-m_\pi^2}(\Theta^p_{\boldsymbol{p}}\Theta^p_{\boldsymbol{k}}+\Theta^n_{\boldsymbol{p}}\Theta^n_{\boldsymbol{k}}+2\Theta^p_{\boldsymbol{p}}\Theta^n_{\boldsymbol{k}})\nonumber\\
    &-\frac{1}{(2\pi)^4}\int \dd |\vec p|\,\dd |\vec k|\,\dd \cos \theta\frac{\vec p^2\vec k^2}{p_0k_0}\frac{C_L^sC_R^s(p_0k_0-|\vec p||\vec k|\cos\theta+m_N^2)}{(q+p-k)^2-m_\pi^2}(\Theta^p_{\boldsymbol{p}}\Theta^p_{\boldsymbol{k}}+\Theta^n_{\boldsymbol{k}}\Theta^n_{\boldsymbol{p}})\nonumber\\
    &+\frac{m_Nq_0}{2(2\pi)^4f^2}\int \dd |\vec p|\,\dd |\vec k|\,\dd \cos \theta\frac{\vec p^2\vec k^2}{p_0k_0}\frac{(p_0+k_0)}{(q+p-k)^2-m_\pi^2}(\Theta^p_{\boldsymbol{p}}\Theta^p_{\boldsymbol{k}}-\Theta^n_{\boldsymbol{k}}\Theta^n_{\boldsymbol{p}})(C_L^s+C_R^s),
\end{align}
\begin{align}
    \Sigma_{\mathrm{2\text{-}loop}}^{0}(q_0)
    &=\frac{m_N^2g_A^2}{10(2\pi f)^4}\int \dd |\vec p|\,\dd |\vec k|\,\dd \cos \theta\frac{\vec p^2\vec k^2}{p_0k_0}\frac{p_0k_0-|\vec p||\vec k|\cos\theta-m_N^2}{(2m_N^2-m_\pi^2-2p_0k_0+2|\vec p||\vec k|\cos\theta)^2}\nonumber\\
    &\times[(-8m_N^2+8p_0k_0-8|\vec p||\vec k|\cos\theta-4q_0^2-2m_\pi^2)(\Theta_{\boldsymbol{k}}^p\Theta^n_{\boldsymbol{p}}+\Theta^p_{\boldsymbol{p}}\Theta^n_{\boldsymbol{k}})\nonumber\\
    &+(-8p_0k_0+8|\vec p||\vec k|\cos\theta+8m_N^2+4q_0^2-3m_\pi^2)(\Theta_{\boldsymbol{k}}^p\Theta_{\boldsymbol{p}}^p+\Theta_{\boldsymbol{k}}^n\Theta_{\boldsymbol{p}}^n)]+\frac{m_N^2g_A^2}{10(2\pi f)^4}\int \dd |\vec p|\,\dd |\vec k|\,\dd \cos \theta\frac{\vec p^2\vec k^2}{p_0k_0}\nonumber\\
    &\times\frac{1}{2m_N^2-m_\pi^2-2p_0k_0+2|\vec p||\vec k|\cos\theta}[(4p_0k_0-4|\vec p||\vec k|\cos\theta-4m_N^2)(\Theta_{\boldsymbol{k}}^p\Theta^n_{\boldsymbol{p}}+\Theta^n_{\boldsymbol{k}}\Theta^p_{\boldsymbol{p}})\nonumber\\
    &+(-4p_0k_0+4|\vec p||\vec k|\cos\theta+4m_N^2)(\Theta^p_{\boldsymbol{p}}\Theta^p_{\boldsymbol{k}}+\Theta^n_{\boldsymbol{p}}\Theta^n_{\boldsymbol{k}})]\nonumber\\
    &-\frac{q_0^2}{2(2\pi)^4f^4}\int \dd |\vec p|\,\dd |\vec k|\,\dd \cos \theta\frac{\vec p^2\vec k^2}{p_0k_0}\frac{p_0k_0+|\vec p||\vec k|\cos\theta+m_N^2}{(q+p-k)^2-m_\pi^2}(\Theta^p_{\boldsymbol{p}}\Theta^p_{\boldsymbol{k}}+\Theta^n_{\boldsymbol{k}}\Theta^n_{\boldsymbol{p}})\nonumber\\
    &-\frac{1}{(2\pi)^4}\int \dd |\vec p|\,\dd |\vec k|\,\dd \cos \theta\frac{\vec p^2\vec k^2}{p_0k_0}\frac{C_L^sC_R^s(p_0k_0-|\vec p||\vec k|\cos\theta+m_N^2)}{(q+p-k)^2-m_\pi^2}(\Theta^p_{\boldsymbol{p}}\Theta^p_{\boldsymbol{k}}+\Theta^n_{\boldsymbol{k}}\Theta^n_{\boldsymbol{p}}),
\end{align}
\end{subequations}
where $C_L^s$ and $C_R^s$ are defined as
\begin{align}
    \MoveEqLeft C_L^s=\frac{8B_0c_1m_q}{f^2}-\frac{2c_2}{f^2m_N^2}q_0p_0(q_0p_0+p_\mu p^\mu-k_\mu p^\mu)\nonumber\\
    &\qquad -\frac{2c_3}{f^2}q_0(q_0+p_0-k_0),\nonumber\\
    \MoveEqLeft C_R^s=\frac{8B_0c_1m_q}{f^2}-\frac{2c_2}{f^2m_N^2}q_0k_0(q_0k_0+k_\mu p^\mu-k_\mu k^\mu)\nonumber\\
    &\qquad -\frac{2c_3}{f^2}q_0(q_0+p_0-k_0).\nonumber
\end{align}
Here, $m_q=(m_u+m_d)/2$ is the isospin-averaged quark mass, $q^\mu = (q^0, \vec 0)$ is the fixed external momentum of the pions, $p^\mu$ and $k^\mu$ are nucleon momenta, $\theta$ is the angle between $\vec p$ and $\vec k$, and $p_0=E(\vec p)=\sqrt{\vec p^2+m_N^2}$. 
Note that some contributions are vanished if $k_F^p = k_F^n$, i.e., when we consider isospin-symmetric nuclear matter. For example, the 7th and 8th diagrams in \cref{self-energy diagrams} are disappeared in this case. In the isospin-asymmetric nuclear matter, we get distinct values of the pion properties for the three pion states. 
After performing these integrals with numerical values, we can compute the in-medium pion masses and the wave function renormalizations by using \cref{eq:pionmass,eq:def-Z}. 
For the derivatives of the self-energies that are needed to compute $Z$, one can either perform the derivatives analytically using the expressions in \cref{eq:result-self-energy}, or perform a numerical derivative by calculating the self-energy twice at $q^\mu$ and $q^\mu+\delta q^\mu$ and dividing by a sufficiently small $\delta q^\mu$. We performed the integrals from the 6th to 9th diagrams in \cref{self-energy diagrams} by taking the leading term in the $1/m_N$ expansion to extract singular parts in the wave function renormalizations, as discussed in \cref{ap:singular}. The other integrals are evaluated numerically. 

The explicit forms of the vertex corrections to the in-medium pion decay constant are given by
\begin{subequations}\label{eq:result-decay-constant}\allowdisplaybreaks
\begin{align}
    \hat{f}_{\mathrm{1\text{-}loop}}^+(q_0)
    &=-\int \dd |\vec p|\frac{g_A^2\vec p^2}{2\pi^2fq_0p_0}\left(\frac{q_0p_0+2\vec p^2}{2p_0+q_0}\Theta^n_{\boldsymbol{p}}+\frac{q_0p_0-2\vec p^2}{2p_0-q_0}\Theta^p_{\boldsymbol{p}}\right)-\int \dd |\vec p|\frac{\vec p^2}{2\pi^2fq_0}(\Theta^p_{\boldsymbol{p}}-\Theta^n_{\boldsymbol{p}})\nonumber\\
    &\qquad +\int \dd |\vec p|\frac{\vec p^2}{\pi^2p_0f}\left(\frac{2c_2}{m_N}p_0^2+2m_Nc_3\right)(\Theta^p_{\boldsymbol{p}}+\Theta^n_{\boldsymbol{p}})\nonumber\\
    &\approx-\frac{m_Ng_A^2}{f(4m_N^2-q_0^2)}\rho+\frac{g_A^2q_0(r-1)}{2f(1+r)(4m_N^2-q_0^2)}\rho+\frac{r-1}{2fq_0(r+1)}\rho+\frac{2(c_2+c_3)}{f}\rho,\\\nonumber\\   
    \hat{f}_{\mathrm{1\text{-}loop}}^-(q_0)
    &=-\int \dd |\vec p|\frac{g_A^2\vec p^2}{2\pi^2fq_0p_0}\left(\frac{q_0p_0+2\vec p^2}{2p_0+q_0}\Theta^p_{\boldsymbol{p}}+\frac{q_0p_0-2\vec p^2}{2p_0-q_0}\Theta^n_{\boldsymbol{p}}\right)+\int \dd |\vec p|\frac{\vec p^2}{2\pi^2fq_0}(\Theta^p_{\boldsymbol{p}}-\Theta^n_{\boldsymbol{p}})\nonumber\\
    &\qquad +\int \dd |\vec p|\frac{\vec p^2}{\pi^2p_0f}\left(\frac{2c_2}{m_N}p_0^2+2m_Nc_3\right)(\Theta^p_{\boldsymbol{p}}+\Theta^n_{\boldsymbol{p}})\nonumber\\
    &\approx-\frac{m_Ng_A^2}{f(4m_N^2-q_0^2)}\rho+\frac{g_A^2q_0(1-r)}{2f(1+r)(4m_N^2-q_0^2)}\rho+\frac{1-r}{2fq_0(1+r)}\rho+\frac{2(c_2+c_3)}{f}\rho,\\\nonumber\\
    \hat{f}_{\mathrm{1\text{-}loop}}^0 (q_0)
    &=-\int \dd |\vec p|\frac{g_A^2\vec p^2}{4\pi^2fq_0p_0}\frac{4p_0^2q_0-4\vec p^2q_0}{4p_0^2-q_0^2}(\Theta^p_{\boldsymbol{p}}+\Theta^n_{\boldsymbol{p}})+\int \dd |\vec p|\frac{2m_N\vec p^2}{\pi^2p_0f}\left(c_2\frac{p_0^2}{m_N^2}+c_3\right)(\Theta^p_{\boldsymbol{p}}+\Theta^n_{\boldsymbol{p}})\nonumber\\
    &\approx-\frac{m_Ng_A^2}{f(4m_N^2-q_0^2)}\rho+\frac{2(c_2+c_3)}{f}\rho,
\end{align}
\begin{align}
    \hat{f}_{\mathrm{2\text{-}loop}}^+(q_0)
    &=\frac{g_A^2m_N^2}{5(2\pi)^4f^3q_0}\int \dd |\vec p|\,\dd |\vec k|\,\dd \cos \theta\frac{\vec p^2\vec k^2}{p_0k_0}\frac{p_0k_0-|\vec p||\vec k|\cos{\theta}-m_N^2}{(2m_N^2-2p_0k_0+2|\vec p||\vec k|\cos{\theta}-m_\pi^2)^2}\nonumber\\
    &\times\left((-2q_0+10p_0-10k_0)\Theta_{\boldsymbol{k}}^p\Theta^n_{\boldsymbol{p}}+(-2q_0-10p_0+10k_0)\Theta^p_{\boldsymbol{p}}\Theta^n_{\boldsymbol{k}}-3q_0(\Theta^p_{\boldsymbol{p}}\Theta^p_{\boldsymbol{k}}+\Theta^n_{\boldsymbol{p}}\Theta^n_{\boldsymbol{k}})\right)\nonumber\\
    &+\frac{g_A^2m_N^2}{4(2\pi)^4f^3q_0}\int \dd |\vec p|\,\dd |\vec k|\,\dd \cos \theta\frac{\vec p^2\vec k^2}{p_0k_0}\frac{k_0-p_0}{2m_N^2-2p_0k_0+2|\vec p||\vec k|\cos{\theta}-m_\pi^2}\nonumber\\
    &\times\left(-4\Theta^p_{\boldsymbol{p}}\Theta^n_{\boldsymbol{k}}+2\Theta^p_{\boldsymbol{k}}\Theta^n_{\boldsymbol{p}}-(\Theta^p_{\boldsymbol{p}}\Theta^p_{\boldsymbol{k}}+\Theta^n_{\boldsymbol{p}}\Theta^n_{\boldsymbol{k}})\right)\nonumber\\
    &+\frac{1}{4(2\pi)^4f^3}\int \dd |\vec p|\,\dd |\vec k|\,\dd \cos \theta\frac{\vec p^2\vec k^2}{p_0k_0}\frac{p_0k_0+|\vec p||\vec k|\cos\theta+m_N^2}{(q+p-k)^2-m_\pi^2}(\Theta^p_{\boldsymbol{p}}\Theta^p_{\boldsymbol{k}}+\Theta^n_{\boldsymbol{k}}\Theta^n_{\boldsymbol{p}}+2\Theta^n_{\boldsymbol{p}}\Theta^p_{\boldsymbol{k}})\nonumber\\
    &-\frac{1}{(2\pi)^4q_0}\int \dd |\vec p|\,\dd |\vec k|\,\dd \cos \theta\frac{\vec p^2\vec k^2}{p_0k_0}\frac{C_L^fC_R^f(p_0k_0-|\vec p||\vec k|\cos\theta+m_N^2)}{(q+p-k)^2-m_\pi^2}(\Theta^p_{\boldsymbol{p}}\Theta^p_{\boldsymbol{k}}+\Theta^n_{\boldsymbol{k}}\Theta^n_{\boldsymbol{p}})\nonumber\\
    &-\frac{m_N}{2(2\pi)^4f^2}\int \dd |\vec p|\,\dd |\vec k|\,\dd \cos \theta\frac{\vec p^2\vec k^2}{p_0k_0}\frac{(p_0+k_0)}{(q+p-k)^2-m_\pi^2}(\Theta^p_{\boldsymbol{p}}\Theta^p_{\boldsymbol{k}}-\Theta^n_{\boldsymbol{k}}\Theta^n_{\boldsymbol{p}})C_R^f\nonumber\\
    &+\frac{m_N}{2(2\pi)^4fq_0}\int \dd |\vec p|\,\dd |\vec k|\,\dd \cos \theta\frac{\vec p^2\vec k^2}{p_0k_0}\frac{(p_0+k_0)}{(q+p-k)^2-m_\pi^2}(\Theta^p_{\boldsymbol{p}}\Theta^p_{\boldsymbol{k}}-\Theta^n_{\boldsymbol{p}}\Theta^n_{\boldsymbol{k}})C_L^f,
\end{align}
\begin{align}
    \hat{f}_{\mathrm{2\text{-}loop}}^-(q_0)
    &=\frac{g_A^2m_N^2}{5(2\pi)^4f^3q_0}\int \dd |\vec p|\,\dd |\vec k|\,\dd \cos \theta\frac{\vec p^2\vec k^2}{p_0k_0}\frac{p_0k_0-|\vec p||\vec k|\cos{\theta}-m_N^2}{(2m_N^2-2p_0k_0+2|\vec p||\vec k|\cos{\theta}-m_\pi^2)^2}\nonumber\\
    &\times\left((-2q_0-10p_0+10k_0)\Theta_{\boldsymbol{k}}^p\Theta^n_{\boldsymbol{p}}+(-2q_0+10p_0-10k_0)\Theta^p_{\boldsymbol{p}}\Theta^n_{\boldsymbol{k}}-3q_0(\Theta^p_{\boldsymbol{p}}\Theta^p_{\boldsymbol{k}}+\Theta^n_{\boldsymbol{p}}\Theta^n_{\boldsymbol{k}})\right)\nonumber\\
    &+\frac{g_A^2m_N^2}{4(2\pi)^4f^3q_0}\int \dd |\vec p|\,\dd |\vec k|\,\dd \cos \theta\frac{\vec p^2\vec k^2}{p_0k_0}\frac{k_0-p_0}{2m_N^2-2p_0k_0+2|\vec p||\vec k|\cos{\theta}-m_\pi^2}\nonumber\\
    &\times\left(2\Theta^p_{\boldsymbol{p}}\Theta^n_{\boldsymbol{k}}-4\Theta^p_{\boldsymbol{k}}\Theta^n_{\boldsymbol{p}}-(\Theta^p_{\boldsymbol{p}}\Theta^p_{\boldsymbol{k}}+\Theta^n_{\boldsymbol{p}}\Theta^n_{\boldsymbol{k}})\right)\nonumber\\
    &+\frac{1}{4(2\pi)^4f^3}\int \dd |\vec p|\,\dd |\vec k|\,\dd \cos \theta\frac{\vec p^2\vec k^2}{p_0k_0}\frac{p_0k_0+|\vec p||\vec k|\cos\theta+m_N^2}{(q+p-k)^2-m_\pi^2}(\Theta^p_{\boldsymbol{p}}\Theta^p_{\boldsymbol{k}}+\Theta^n_{\boldsymbol{k}}\Theta^n_{\boldsymbol{p}}+2\Theta^p_{\boldsymbol{p}}\Theta^n_{\boldsymbol{k}})\nonumber\\
    &-\frac{1}{(2\pi)^4q_0}\int \dd |\vec p|\,\dd |\vec k|\,\dd \cos \theta\frac{\vec p^2\vec k^2}{p_0k_0}\frac{C_L^fC_R^f(p_0k_0-|\vec p||\vec k|\cos\theta+m_N^2)}{(q+p-k)^2-m_\pi^2}(\Theta^p_{\boldsymbol{p}}\Theta^p_{\boldsymbol{k}}+\Theta^n_{\boldsymbol{k}}\Theta^n_{\boldsymbol{p}})\nonumber\\
    &+\frac{m_N}{2(2\pi)^4f^2}\int \dd |\vec p|\,\dd |\vec k|\,\dd \cos \theta\frac{\vec p^2\vec k^2}{p_0k_0}\frac{(p_0+k_0)}{(q+p-k)^2-m_\pi^2}(\Theta^p_{\boldsymbol{p}}\Theta^p_{\boldsymbol{k}}-\Theta^n_{\boldsymbol{k}}\Theta^n_{\boldsymbol{p}})C_R^f\nonumber\\
    &-\frac{m_N}{2(2\pi)^4fq_0}\int \dd |\vec p|\,\dd |\vec k|\,\dd \cos \theta\frac{\vec p^2\vec k^2}{p_0k_0}\frac{(p_0+k_0)}{(q+p-k)^2-m_\pi^2}(\Theta^p_{\boldsymbol{p}}\Theta^p_{\boldsymbol{k}}-\Theta^n_{\boldsymbol{k}}\Theta^n_{\boldsymbol{p}})C_L^f,
\end{align}
\begin{align}
    \hat{f}_{\mathrm{2\text{-}loop}}^0(q_0)
    &=\frac{g_A^2m_N^2}{5(2\pi)^4f^3q_0}\int \dd |\vec p|\,\dd |\vec k|\,\dd \cos \theta\frac{\vec p^2\vec k^2}{p_0k_0}\frac{p_0k_0-|\vec p||\vec k|\cos{\theta}-m_N^2}{(2m_N^2-2p_0k_0+2|\vec p||\vec k|\cos{\theta}-m_\pi^2)^2}\nonumber\\
    &\times\left(-6q_0\Theta_{\boldsymbol{k}}^p\Theta^n_{\boldsymbol{p}}-6q_0\Theta^p_{\boldsymbol{p}}\Theta^n_{\boldsymbol{k}}+q_0(\Theta^p_{\boldsymbol{p}}\Theta^p_{\boldsymbol{k}}+\Theta^n_{\boldsymbol{p}}\Theta^n_{\boldsymbol{k}})\right)-\frac{g_A^2m_N^2}{2(2\pi)^4f^3q_0}\int \dd |\vec p|\,\dd |\vec k|\,\dd \cos \theta\frac{\vec p^2\vec k^2}{p_0k_0}\nonumber\\
    &\times\frac{k_0-p_0}{2m_N^2-2p_0k_0+2|\vec p||\vec k|\cos{\theta}-m_\pi^2}\left(\Theta^p_{\boldsymbol{p}}\Theta^n_{\boldsymbol{k}}+\Theta^p_{\boldsymbol{k}}\Theta^n_{\boldsymbol{p}}\right)\nonumber\\
    &+\frac{1}{2(2\pi)^4f^3}\int \dd |\vec p|\,\dd |\vec k|\,\dd \cos \theta\frac{\vec p^2\vec k^2}{p_0k_0}\frac{p_0k_0+|\vec p||\vec k|\cos\theta+m_N^2}{(q+p-k)^2-m_\pi^2}(\Theta^p_{\boldsymbol{p}}\Theta^n_{\boldsymbol{k}}+\Theta^n_{\boldsymbol{p}}\Theta^p_{\boldsymbol{k}})\nonumber\\
    &-\frac{1}{(2\pi)^4q_0}\int \dd |\vec p|\,\dd |\vec k|\,\dd \cos \theta\frac{\vec p^2\vec k^2}{p_0k_0}\frac{C_L^fC_R^f(p_0k_0-|\vec p||\vec k|\cos\theta+m_N^2)}{(q+p-k)^2-m_\pi^2}(\Theta^p_{\boldsymbol{p}}\Theta^p_{\boldsymbol{k}}+\Theta^n_{\boldsymbol{k}}\Theta^n_{\boldsymbol{p}}),
\end{align}
\end{subequations}
where $C_L^f$ and $C_R^f$ are written as
\begin{align}
    \MoveEqLeft C_L^f=\frac{8B_0c_1m_q}{f^2}-\frac{2c_2}{f^2m_N^2}q_0p_0(q_0p_0+p_\mu p^\mu-k_\mu p^\mu)\nonumber\nonumber\\
    &\qquad -\frac{2c_3}{f^2}q_0(q_0+p_0-k_0), \nonumber\\
    \MoveEqLeft C_R^f=\frac{2c_2}{fm_N^2}k_0(q_0k_0+k_\mu p^\mu-k_\mu k^\mu)+\frac{2c_3}{f}(q_0+p_0-k_0).\nonumber
\end{align}
\end{widetext}
Again, we show the linear contributions of nuclear density for the one-loop corrections on the right-hand side of $\approx$. The integrals are evaluated in a similar way to self-energies. After performing these integrals with numerical values, we first compute $\hat f_{\pi^{\pm,0}}$ and then multiply the square root of the corresponding wave function renormalization, according to $f^*_{\pi^{\pm,0}} = \hat f_{\pi^{\pm,0}} \sqrt{Z_{\pi^{\pm,0}}}$. 

\section{Singularities in wave function renormalizations}\label{ap:singular}
In the calculation of the wave function renormalizations, we obtain some divergence terms. From the 6th and 9th diagrams in ~\cref{self-energy diagrams,axial 1PI diagrams} at the next-to-leading order, we get integrals of the following form:
\begin{widetext}
\begin{align}
    &\int \dd[3] \vec p \, \dd[3] \vec k \frac{1}{2m_N^2-2k_0p_0+2pk\cos\theta+2q_0(p_0-k_0)+q_0^2-m_\pi^2}\Theta(k_F^{p,n}-|\boldsymbol{p|)}\Theta(k_F^{p,n}-|\boldsymbol{k)|}
\end{align}
We change the momentum convention $\boldsymbol{p_i}\rightarrow\boldsymbol{k_i}-\frac{\boldsymbol{p_i}}{2}, \boldsymbol{k_i}\rightarrow\boldsymbol{k_i}+\frac{\boldsymbol{p_i}}{2}$ to perform the integral of the Heaviside step functions geometrically, and take the leading order of large-$m_N$ limit, which gives $\mathcal{O}(\rho^{4/3})$ corrections. From the next-to-leading order in the large-$m_N$ limit, it provides higher-order terms than our purpose. Then, the integral changes to
\begin{align}
    &\int \dd[3] \vec p \, \dd[3] \vec k \frac{1}{-p^2+q_0^2-m_\pi^2}\Theta\left(k_F^{p,n}-\left|\boldsymbol{k}-\frac{\boldsymbol{p}}{2}\right|\right)\Theta\left(k_F^{p,n}-\left|\boldsymbol{k}+\frac{\boldsymbol{p}}{2}\right|\right)
\label{asing}
\end{align}
This integral gives finite contributions in the limit $q_0\rightarrow m_\pi$ without singularity. However, when we take derivatives \cref{asing} with respect to $q_0$, we get
\begin{align}
    &\int \dd[3] \vec p \, \dd[3] \vec k \frac{-2q_0}{(p^2-q_0^2+m_\pi^2)^2}\Theta\left(k_F^{p,n}-\left|\boldsymbol{k}-\frac{\boldsymbol{p}}{2}\right|\right)\Theta\left(k_F^{p,n}-\left|\boldsymbol{k}+\frac{\boldsymbol{p}}{2}\right|\right).
\end{align}
\end{widetext}
This integral diverges as $\boldsymbol{p}\rightarrow0$ in the limit $q_0\rightarrow m_\pi$. This singularity is analogous to infrared singularities in quantum field theory, and it is expected that those singular parts in the corrections are canceled by additional real emission diagrams when obtaining relevant scattering cross-sections. Therefore, we abandon singular parts from integrals with similar forms in the wave function renormalizations, which are related to the two-point correlation functions of the pions. More detailed arguments can be read in Ref.~\cite{Goda2014}.

\bibliographystyle{apsrev4-2}
\bibliography{references.bib}

\end{document}